\documentclass[aps,prb,twocolumn,showpacs,showkeys,footinbib,superscriptaddress]{revtex4-1}

\usepackage{amsmath}
\usepackage{amssymb}
\usepackage{graphicx}
\usepackage{bm}
\usepackage{color}
\usepackage{relsize}
\usepackage{braket}
\usepackage[caption=false]{subfig}
\usepackage{enumerate} 
\usepackage[makeroom]{cancel}
\usepackage{hyperref}
\usepackage[all]{hypcap}

\begin{document}
\title{Localization-induced optical properties of monolayer transition-metal dichalcogenides}
\date{\today}

\author{Dinh Van Tuan}
\email[]{vdinh@ur.rochester.edu}
\affiliation{Department of Electrical and Computer Engineering, University of Rochester, Rochester, New York 14627, USA}
\author{Hanan~Dery}
\email[]{hanan.dery@rochester.edu}
\affiliation{Department of Electrical and Computer Engineering, University of Rochester, Rochester, New York 14627, USA}
\affiliation{Department of Physics and Astronomy, University of Rochester, Rochester, New York 14627, USA}

\begin{abstract} 
Impurities play an important role during recombination processes in semiconductors. Their important role is sharpened in atomically-thin transition-metal dichalcogenides whose two-dimensional character renders electrons and holes highly susceptible to localization caused by remote charged impurities.  We study a multitude of phenomena that arise from the interaction of localized electrons with excitonic complexes. Emphasis is given to the amplification of the phonon-assisted recombination of biexcitons when it is mediated by localized electrons, showing that this mechanism can explain recent photoluminescence experiments in ML-WSe$_2$. In addition, the magnetic-field dependence of this mechanism is analyzed.  The results of this work point to (i) an intriguing coupling between the longitudinal-optical and homopolar phonon modes that can further elucidate various experimental results, (ii) the physics behind a series of localization-induced optical transitions in tungsten-based materials, and (iii) the importance of localization centers in facilitating the creation of biexcitons and exciton-exciton annihilation processes.   
\end{abstract}

\pacs{}
\keywords{}

\maketitle

\section{Introduction}

Monolayer transition-metal dichalcogenides (ML-TMDs) are promising semiconductors  for ultrathin photonic and optoelectronic devices.\cite{Wang_NatNano12,Britnell_Science13,Sundaram_NanoLett13,Ross_NatNano14,Xu_NatPhys14,Mak_NatPhot16,Wang_RMP18} An advantage of these van der Waals layered materials is their compatibility with flexible substrates and various heterostructures.\cite{Geim_Nature13} For example, stamping ML-TMDs on silicon crystals  can alleviate the long-standing problem of finding direct-gap semiconductors that are lattice matched to silicon, thereby paving the way for new optics-on-chip architectures. Owing to the direct band-gap energy and two-dimensional (2D) nature of ML-TMDs,\cite{Splendiani_NanoLett10,Mak_PRL10,Korn_APL11} the optical absorbance in these atomic monolayers is markedly strong and governed by creation of tightly bound excitons (electron-hole pairs).\cite{Chernikov_PRL14,He_PRL14,Stier_PRL18,Liu_arXiv18} The exciton binding energy can be an order of magnitude larger than the room-temperature thermal energy when an undoped monolayer is embedded in or supported on materials with low dielectric constants.\cite{Berkelbach2013:PRB,Thilagam_JAP14,Wang_ADP14,Berghauser_PRB15,Zhang_NanoLett15,Ganchev_PRL15,Mayers_PRB15,Velizhanin_PRB15,Latini_PRB15,Qiu_PRB16,Kidd_PRB16,Borghardt_PRM17,Trolle_SR17,Mostaani_PRB17,Raja_NatComm17,Meckbach_PRB18,VanTuan_PRB18}  

In addition to being tightly bound, the exciton states in ML-TMDs are embellished by the coupling between the spin and valley (pseudo-spin) degrees of freedom.\cite{Xiao_PRL12,Xiao_PRL12,Zeng_NatNano12,Mak_NatNano12,Feng_NatComm12,Jones_NatNano13,Song_PRL13,Liu_PRB13,Zhou_PRL15,Bieniek_PRB18,Trushin_PRL18} Because ML-TMDs are non centro-symmetric crystals, time-reversal partners the spin and valley quantum numbers such that spin-up charge carriers populate valleys (low-energy pockets) on one side of the Brillouin zone, while spin-down ones populate valleys on the opposite side.\cite{Xiao_PRL12,Song_PRL13}  The combination of these quantum numbers facilitates a miscellany of few-body complexes, which can be probed in photoluminescence (PL) experiments on accounts of the small spin-splitting energy at the edge of the conduction band compared with the exciton binding energy.\cite{Song_PRL13,Dery_PRB15} These include two-body complexes such as direct, indirect, and dark excitons,\cite{Song_PRL13,Dery_PRB15,Wu_PRB15,Qiu_PRL15,Zhang_PRL15,Slobodeniuk_2DMater16,Echeverry_PRB16,Selig_NatCommun16,Selig_2DMater18,Malic_PRM18,Lindlau_arXiv17,Robert_PRB17,Zhang_NatNano17,Zhou_NatNano17,Wang_PRL17,Scharf_PRL17,Molas_2DMater17,Molas_arXiv19} three-body complexes such as intravalley and intervalley trions,\cite{Jones_NatPhys16,Plechinger_NatCom16,Plechinger_NanoLett16,Courtade_PRB17,Wang_NanoLett17,Wang_NatNano17} as well as four- and five-body complexes such as neutral and charged biexcitons.\cite{Hao_NatCommun17,Chen_NatComm18,Ye_NatComm18,Li_NatComm18,Barbone_NatComm18} In addition, shakeup processes due to the coupling of excitons to collective charge excitations when electrons transition between time-reversed valleys lead to optical sidebands.\cite{Dery_PRB16,VanTuan_PRX17,VanTuan_PRB19,Scharf_JPCM19}

The progress made in the fabrication of high-quality devices has sharpened the understanding of emission processes in ML-TMDs. Prior to the time that encapsulating ML-TMDs in hexagonal boron nitride (hBN) became a common practice, the emission spectrum from relatively low-quality MLs supported on quartz substrates had typically shown two broad peaks.\cite{VanTuan_arXiv19} The high-energy one was attributed to neutral excitons and the low-energy peak  to negative trions because the MLs were unintentionally electron doped. With the advance in fabrication techniques,\cite{Wang_Science13,CastellanosGomez_2DMater14,Amani_Science15,Cadiz_PRX17} not only the spectral lines became narrower and the fine-structure of trion states was revealed, but the binding energy of the trion when extracted from the emission spectrum was different.  For example, early PL experiments in ML-WS$_2$ supported on SiO$_2$ have reported the trion peak to be $\sim$43~meV below the neutral exciton.\cite{Mitioglu_PRB13,Plechinger_PSS15} Recent experiments with higher-quality ML-WS$_2$/SiO$_2$ devices, on the other hand, have shown that the peaks attributed to the fine-structure of negative trions emerge closer to the neutral exciton,\cite{NaglerPRL18} whereas the peak that emerges $\sim$43~meV below the neutral exciton is phonon-assisted recombination.\cite{Plechinger_NanoLett16} Indeed, the energy of the longitudinal-optical (LO) phonon mode in ML-WS$_2$ seen in Raman experiments and calculated in DFT calculations is 44~meV.\cite{VanTuan_arXiv19,MolinaSanchez_PRB11}

Recently, we have shown that phonon-assisted recombination of an exciton is a strong process when the exciton interacts with a localized electron.\cite{VanTuan_arXiv19} The recombination process involves emission of an optical phonon when the localized electron captures the exciton to form an intermediate virtual trion state, followed by emission of a photon. The emission process is strong if the energy level of the intermediate state is close to that of a real trion. In addition, we have found that the interaction between the exciton and the electron is effective when the localization of the latter is relatively tight and when the initial kinetic energy of the exciton is of the order of few meV or less.\cite{VanTuan_arXiv19} These findings suggest that the formation of trions is more effective next to localization centers compared with the case that their constituents are initially itinerant (delocalized electron and exciton). Further evidence for this fact is corroborated in several recent experiments that have studied the PL through two-dimensional hyperspectral mapping.\cite{Bao_NatComm15,Neumann_PRM18} Time-integrated and time-resolved PL experiments at various temperatures also point to the important role of localization during emission processes in ML-TMDs.\cite{Godde_PRB16,Jadczak_nanotech17}

The goal of this work is to further unfold the interplay between localization and optical transitions in ML-TMDs. Implications of loosing the valley degree of freedom are addressed first. Following this discussion, we analyze the findings of recent PL experiments that detect five-particle complexes in the emission spectrum of ML-WSe$_2$ devices.\cite{Chen_NatComm18,Ye_NatComm18,Li_NatComm18,Barbone_NatComm18} The phonon-assisted recombination of biexcitons next to localized electrons emerges as a strong candidate to explain the experimental results in a consistent way. Relying solely on empirical evidence, we find that the spectral position of the five-particle peak is what one should expect from the phonon-assisted recombination process. We then quantify the dependence of this process on the amplitude of an out-of-plane magnetic field, finding agreement with experiment. Localization effects are shown to manifest in the magnetic-field dependent PL through amplification and suppression of different optical transitions. 

Throughout this work, we point to various phenomena that can shed light on a wide range of observations seen in PL experiments. For example, we suggest an intriguing coupling between the homopolar and LO phonons in ML-TMDs that can explain why experiments show that the phonon with lower energy of these two modes dominates the Raman spectrum as well as the phonon-assisted recombination process. Specifically, the homopolar phonon dominates in selenide-based MLs and the LO phonon in sulfides-based MLs.\cite{VanTuan_arXiv19} In addition, we elucidate the following localization-related aspects: (i) the sub-quadratic amplitude dependence of the five-particle peak on excitation light intensity, (ii) how to tell apart recombination processes that involve localized and delocalized trions, (iii) how to recognize the involvement of phonons in the recombination process, and (iv) how indirect or dark excitons turn to optically-active localized trions. Finally, we discuss the importance of localization centers in facilitating the creation of biexcitons or exciton-exciton annihilation processes. 

This paper is organized as follows. Section~\ref{sec:localization} deals with the type of electron localization studied in this work. Section~\ref{sec:energy} describes the experimental findings for neutral and charged biexcitons, where we analyze two possible interpretations for these experiments. Section~\ref{sec:theory} includes the theory of the phonon-assisted recombination process where we study the capture mechanism and dependence of the recombination on magnetic field. In addition, we point to the coupling between homopolar and LO phonon modes. Section~\ref{sec:discussion} is a comprehensive discussion of various aspects of the interplay between localization and recombination processes. We conclude our findings in Sec.~\ref{sec:conclusions} and provide an outlook for future studies and research directions. Appendix~\ref{app:V} provides details on the Coulomb interaction between charged particles in the ML and between these particles and a remote charged impurity in the substrate.  Appendix~\ref{app:SVM} provides technical details on the computation methods that we use to calculate the few-body states, Fr\"{o}hlich interaction, and the matrix element of the capture process. Appendix~\ref{app:parameters} lists the parameter values that we use in our simulations, and Appendix~\ref{app:benchmark_delocalized} benchmarks the calculated energies of few-body complexes against experiment. 

\section{Electron localization}\label{sec:localization}

The relatively large surface-area-to-volume ratio of 2D atomic crystals renders them highly susceptible to impurities in the surrounding dielectric environment compared with bulk crystals. That is, even when dealing with pristine 2D crystals, their transport and optical properties are still affected by charged impurities in the substrate, or admolecules/adatoms at its surface.  One famous related example is the electron-hole puddles in graphene giving rise to its anomalous non-zero minimal conductivity at zero average carrier density.\cite{AdamPRB2011,Jung_NatPhys11,Zhang_NatPhys09,Deshpande_PRB09} Below, we first describe the type of localization we consider in this work, followed by discussion of the valley mixing of a localized electron and the type of delocalized few-body complexes that can interact with it. We then discuss the changes needed when dealing with localized hole states. All of these aspects will be instrumental for the analysis of experimental results in the following sections.  

\begin{figure}
 \centering
\includegraphics[width=8.5 cm]{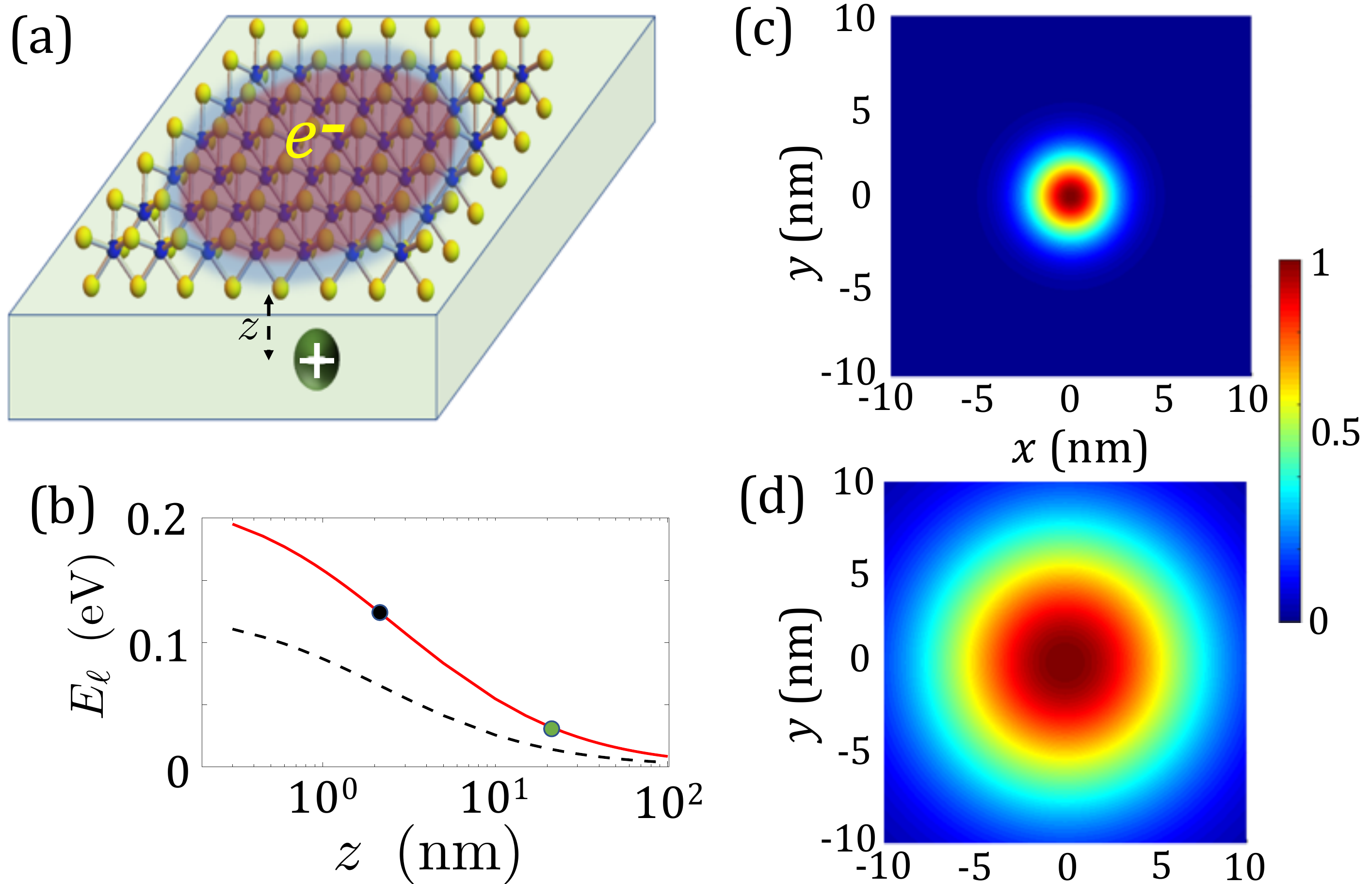}
 \caption{(a) A cartoon of a localization center where an electron in the monolayer is localized in the vicinity of a positively-charged defect in the substrate. (b) Binding energy of a localized electron in ML-WSe$_2$ as a function of the distance between the monolayer and a positively charged defect. The solid  line shows a supported ML where the defect is embedded in SiO$_2$, and the dashed line corresponds to encapsulated ML where the defect is embedded in hBN. (c) and (d) Normalized spatial distributions of the localized-electron ground state, $\left| \Psi_\ell(x,y) \right|^2$,  in the supported ML where the charged defect coordinates are (0,0,$z=2$~nm) and (0,0,$z=20$~nm), respectively. The corresponding binding energies are denoted by the black and green dots in (b).}\label{fig:localE}
\end{figure}

\begin{figure*}
 \centering
\includegraphics[width=16 cm]{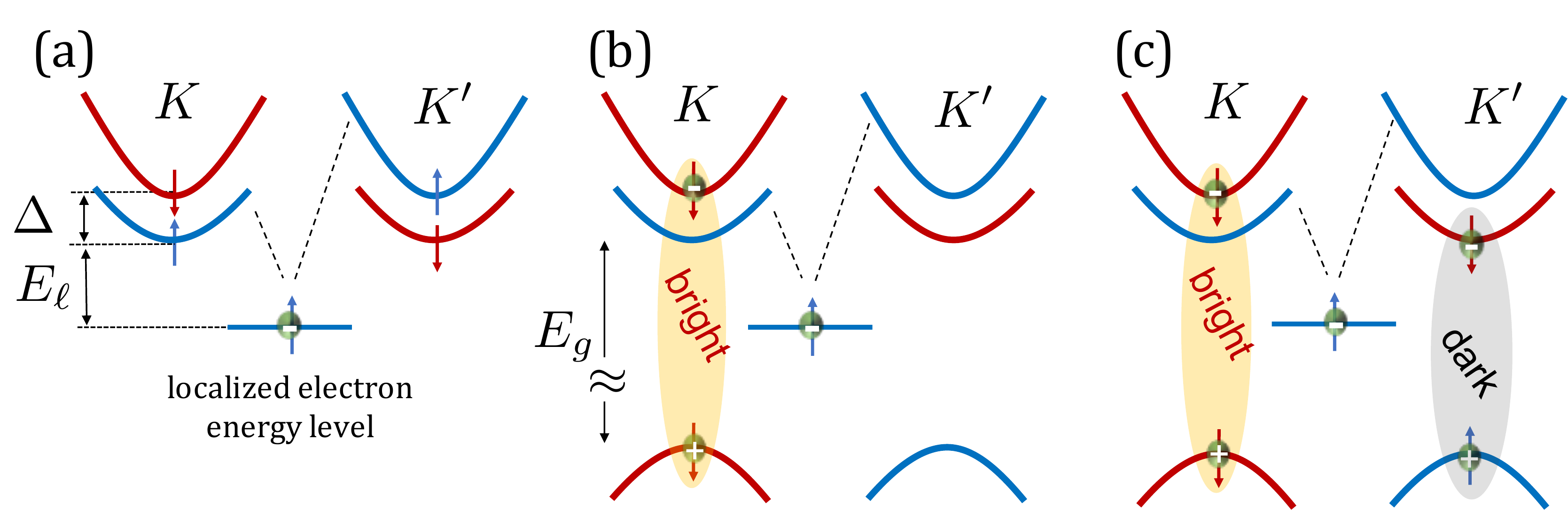}
 \caption{The state of a spin-up  localized electron and the excitonic complexes that it interacts with.  (a)  The state of the localized electron is a superposition of conduction-band edge states with the same spin. Its binding energy is comparable or larger than the conduction-band spin splitting energy but much smaller than the band-gap energy ($E_g \gg E_\ell \gtrsim \Delta$).  (b) The initial state of the phonon-assisted recombination process of excitons in ML-WSe$_2$ (or ML-WS$_2$). (c) The same but for biexcitons. The state of a spin-down localized electron and its associated complexes are the time-reversed analogues of those shown in (a)-(c).}\label{fig:diagram}
\end{figure*}

\subsection{Localization due to remote charged impurities}

The type of localization we allude to in this work is shown by the cartoon in Fig.~\ref{fig:localE}(a), where an electron in the ML  is localized to the vicinity of a positively charged impurity which is embedded in the substrate. Figure~\ref{fig:localE}(b) shows the binding energies of an electron in ML-WSe$_2$ supported on SiO$_2$ (solid line) and encapsulated in hBN (dashed line) as a function of the distance of the charged impurity from the ML. Appendices~\ref{app:V}, \ref{app:SVM} and \ref{app:parameters} include details on the Coulomb potential, Wannier-Mott model we have used to calculate the localized states, and employed parameters, respectively. Figures~\ref{fig:localE}(c) and (d) show the squared amplitude of the electron's ground-state wavefunction for the case that the impurity is 2 and 20~nm away from the mid-plane of a supported ML. 

The Bloch waves of electrons and holes in ML-TMDs are governed by the $d$-orbitals of the transition-metal atoms in the mid-plane of the ML.\cite{Zhu_PRB11,Xiao_PRL12, Kormanyos_2DMater15} The lack of overlap between these orbitals and the remote impurity leads to a weak influence on neutral complexes such as excitons or biexcitons. For example, while the binding energy of an electron to a surface charge impurity can be larger than 100~meV, as shown in Fig.~\ref{fig:localE}(b), the binding of the exciton to such defect is only $\sim$1~meV.\cite{VanTuan_arXiv19}  On the other hand, the interaction between the electron that the impurity strongly localizes and the particles of the neutral complex is strong because they contact each other (i.e., their wavefunctions overlap).  

\subsection{Valley mixing}

The wavefunction of a shallow localized state in a semiconductor is approximated by the product of an envelope function and periodic function. This approximation assumes that the localization length is much larger than the lattice constant, or equivalently, that the electron binding energy is much smaller than the band-gap energy ($E_g \gg E_{\ell}$ where $E_g$ is of the order of 2~eV in ML-TMDs). If this condition is met then one can employ a Mott-Wannier model to calculate the binding energies and envelope functions, as we did in Fig.~\ref{fig:localE}. 

The periodic part of the wavefunction of a localized electron in multivalley semiconductors comes from a superposition of the edge states in the conduction band when describing shallow localized states.\cite{Kohn_PR55,Kohn_SSP57,Song_PRL14} For example, the periodic function part of donor states in Si is described through superposition of its six conduction-band minima  states, or in Ge through superposition of its four conduction-band minima  states.\cite{Kohn_PR55,Kohn_SSP57}  Equivalently, the periodic function of a shallow localized electron state in ML-TMDs comprises components from the conduction-band edge states at the $K$ and $K'$ points, as shown in Fig.~\ref{fig:diagram}(a). When the binding energy of the localized electron is not negligible compared with the spin-spitting energy of the conduction band, $E_{\ell} \not \ll \Delta$, then both edge states contribute measurably to the periodic function. Consequently, the valley degree of freedom is no longer a good quantum number in this case.  Given that $\Delta$ is 0.03~eV or less in ML-TMDs,\cite{Kosmider_PRB13,Cheiwchanchamnangij_PRB13,Kormanyos_2DMater15} one can infer from Fig.~\ref{fig:localE}(b) that the valley mixing of localized electron states should be evident unless they are weakly bound to a remote charged impurity whose distance from the ML is more than a few tens nm. 

\subsection{Implications of the Pauli-exclusion principle} \label{sec:pauli}
Considering the valley mixing effect,  the Pauli-exclusion principle dictates that a photoexcited bright exciton interacts with the localized electron when the spins of the involved electrons are opposite. Figure~\ref{fig:diagram}(b) shows a cartoon of such exciton and spin-up localized electron. The phonon-assisted recombination of an exciton when mediated by a localized electron takes place when the latter captures the exciton by emitting a phonon, followed by photon emission from the intermediate virtual trion state.\cite{VanTuan_arXiv19} This process has one emission peak, governed by a singlet spin configuration of the two electrons. Yet, the intermediate virtual trion is not an intravalley trion because of the valley-mixed state that characterizes the localized electron.  The picture is different for trions or other few-body complexes that are free to move in the monolayer. In this case, the valley degree of freedom remains a good quantum number for each particle in the few-body complex. As a result, we can have intravalley and intervalley delocalized trions in the emission or absorption spectra (with singlet spin configuration when electrons come from the same valley or triplet spin configuration when electrons come from opposite valleys).\cite{Jones_NatPhys16,Plechinger_NanoLett16,Courtade_PRB17,Wang_NanoLett17}

Similarly, Fig.~\ref{fig:diagram}(c) shows the biexciton that can interact with the spin-up localized electron. Here, the only way for the three involved electrons to have different quantum numbers is if the delocalized biexciton comprises dark and bright exciton components whose electrons come from opposite valleys but have the same spin. \cite{Chen_NatComm18,Ye_NatComm18,Li_NatComm18,Barbone_NatComm18} This spin is opposite to that of the valley-mixed localized electron, as shown in Fig.~\ref{fig:diagram}(c).  The phonon-assisted recombination of a biexciton when mediated by a localized electron takes place when the latter captures the biexciton by emitting a phonon, followed by photon emission from the intermediate virtual five-particle state. We will elaborate on this mechanism in Sec.~\ref{sec:exp_5_phonon}. 

\subsection{Localized holes} \label{sec:holes}
When dealing with shallow localized hole states in ML-TMDs due to negatively-charged remote impurities, the valley degree of freedom may still be a good quantum number because of the relatively large spin-splitting energy in the valence band ($\sim$$200$~meV in MoS$_2$ or MoSe$_2$,  $\sim$$ 400$~meV in WS$_2$ or WSe$_2$).\cite{Zhu_PRB11,Xiao_PRL12, Kormanyos_2DMater15} That is, the periodic function of the localized state is predominantly governed by the top valley of the valence band with the same spin direction. As a result, the phonon-assisted recombination of a biexciton cannot be mediated by a localized hole because there is no way for three holes to have different quantum numbers when they are governed by states of the valence-band top valleys. This description is consistent with the empirical finding that the five-particle complex involves three electrons and two holes.\cite{Chen_NatComm18,Ye_NatComm18,Li_NatComm18,Barbone_NatComm18} On the other hand, localized holes can mediate the phonon-assisted recombination of excitons for which the process involves only two holes. However, the recombination can be strong only when the phonon energy is close to the binding energy of the positive trion to ensure a resonance condition. 

\section{Biexcitons and five-particle complexes} \label{sec:energy}
The experiments in Refs.~[\onlinecite{Chen_NatComm18,Ye_NatComm18,Li_NatComm18,Barbone_NatComm18}] have studied the PL in hBN-encapsulated ML-WSe$_2$ as a function of excitation light intensity, gate voltage, and magnetic field. In agreement with previous findings,\cite{Jones_NatNano13,Courtade_PRB17,Wang_NanoLett17,Jones_NatPhys16} (a) Delocalized bright and dark excitons were identified when the gate-induced electrostatic  doping was small, (b) The delocalized positive trion was identified when holes were added to the ML at negative gate voltages, (c)  Delocalized intravalley and intervalley negative trions were identified when electrons were added to the ML at intermediate positive gate voltages, and (d) the strong emission from the low-energy part of the spectrum was detected at elevated electron densities (large positive gate voltages). We have attributed the latter to the shakeup process when excitons couple to shortwave collective charge excitations (intervalley plasmons).\cite{Dery_PRB16,VanTuan_PRX17,VanTuan_PRB19,Scharf_JPCM19} Compared with previous experiments,\cite{Courtade_PRB17,Wang_NanoLett17,VanTuan_PRB19,Jones_NatPhys16} all of these peaks emerge in their expected spectral positions and they show linear dependence on the excitation light intensity.


\begin{table} [t!]
\renewcommand{\arraystretch}{1.25}
\tabcolsep=0.10 cm
\caption{\label{tab:EPosition} The photon energies of the direct neutral exciton ($X^0$),  neutral biexciton ($XX^0$),  and negatively charged biexciton ($XX^-$). The data is extracted from gate-voltage-dependent PL in Refs.~[\onlinecite{Chen_NatComm18,Ye_NatComm18,Li_NatComm18,Barbone_NatComm18}].  The energy unit is meV.   $\mathcal{E}_{XX^0}$ is the biexciton binding energy (difference between the 1$^{\text{st}}$ and 2$^{\text{nd}}$ columns). $\Delta_{XX}$ is the energy difference between the peaks of the neutral and charged biexcitons (2$^{\text{nd}}$ and 3$^{\text{rd}}$ columns).}
\begin{tabular}{|c|ccc||cc|}
\hline\hline
Ref. 						&  \,\,\,$\hbar\omega_{X^0}$\,\,\, 	&  \,\,\,$\hbar\omega_{XX^0}$\,\,\,    	& \,\,\,$\hbar\omega_{XX^-}$\,\,\,	& \,\,\,\,\,\,$\mathcal{E}_{XX^0}$\,\,\,\,\,\, 	& \,\,\,\,\,\,$\Delta_{XX}$\,\,\,\,\,\,    \\ \hline
\onlinecite{Chen_NatComm18}  	& 1730 			& 1712   		& 1681   		    				&      18						& 31		  \\ \hline
\onlinecite{Ye_NatComm18}  		& 1723			& 1703  		& 1671   		    				&	20						& 32		  \\ \hline
\onlinecite{Li_NatComm18}  		& 1724 			& 1708   		& 1676   						&	16						& 32		  \\ \hline	
\onlinecite{Barbone_NatComm18}  	& 1726 			& 1709   		& 1677   						&	17						& 32	    	   \\ 
\hline \hline
\end{tabular}
\vspace{-2mm}
\end{table}

The unique experimental findings are the identifications of the biexciton and five-particle complex.\cite{Chen_NatComm18,Ye_NatComm18,Li_NatComm18,Barbone_NatComm18}  Table \ref{tab:EPosition} lists the detected photon energies of the neutral and charged biexcitons as well as of the bright exciton. The neutral biexciton peak, $XX^0$, emerged between 16 and 20 meV below the bright exciton peak, in agreement with theoretical models.\cite{Kylanpaa_PRB15,Kezerashvili_FBS17,Szyniszewski_PRB17,Donck_PRB17,Kezerashvili_arXvi19} Its dependence on excitation light intensity was quadratic as expected, and the magnetic field analysis revealed that it comprises bright and dark exciton components [the diagram in Fig.~\ref{fig:diagram}(c) but without the localized electron]. The peak of the five-particle complex (negatively charged biexciton), which we will denote by $XX^-$, emerged between 48 and 52 meV below the bright exciton peak. Its  dependence on excitation light intensity varied in these experiments, but all show a sub-quadratic power law. 

The peak $XX^-$ was previously detected in ML-WSe$_2$ supported on SiO$_2$,\cite{You_NatPhys15} where its original interpretation was not conclusive because of the lack of gate-voltage control. For example, we have erroneously interpreted this peak as a biexciton whose binding energy is governed by the shortwave Coulomb interaction.\cite{VanTuan_PRX17} We have assumed so because neutral biexcitons have a binding energy in the ballpark of 20~meV,\cite{Kylanpaa_PRB15,Kezerashvili_FBS17,Szyniszewski_PRB17,Donck_PRB17,Kezerashvili_arXvi19} whereas the binding energy found in Ref.~[\onlinecite{You_NatPhys15}] was 52~meV. In addition, the peak emerged in close proximity to the optical sideband, which is seen in other experiments and attributed to coupling between exciton and intervalley plasmons.\cite{VanTuan_PRB19,Wang_NanoLett17,Jones_NatNano13} However, the findings in Refs.~[\onlinecite{Chen_NatComm18,Ye_NatComm18,Li_NatComm18,Barbone_NatComm18}] rule out this explanation. They clearly show that the peak $XX^-$ appears in the emission spectrum only at relatively small electron densities, and it disappears at large electron densities in which plasmon effects can actually play a role. Indeed, the plasmon-related optical sideband emerges at larger electron densities and at a slightly different energy.\cite{Jones_NatNano13,Dery_PRB16,Chen_NatComm18,Ye_NatComm18,Li_NatComm18,Barbone_NatComm18}

\begin{figure}
 \centering
\includegraphics[width=8.5 cm]{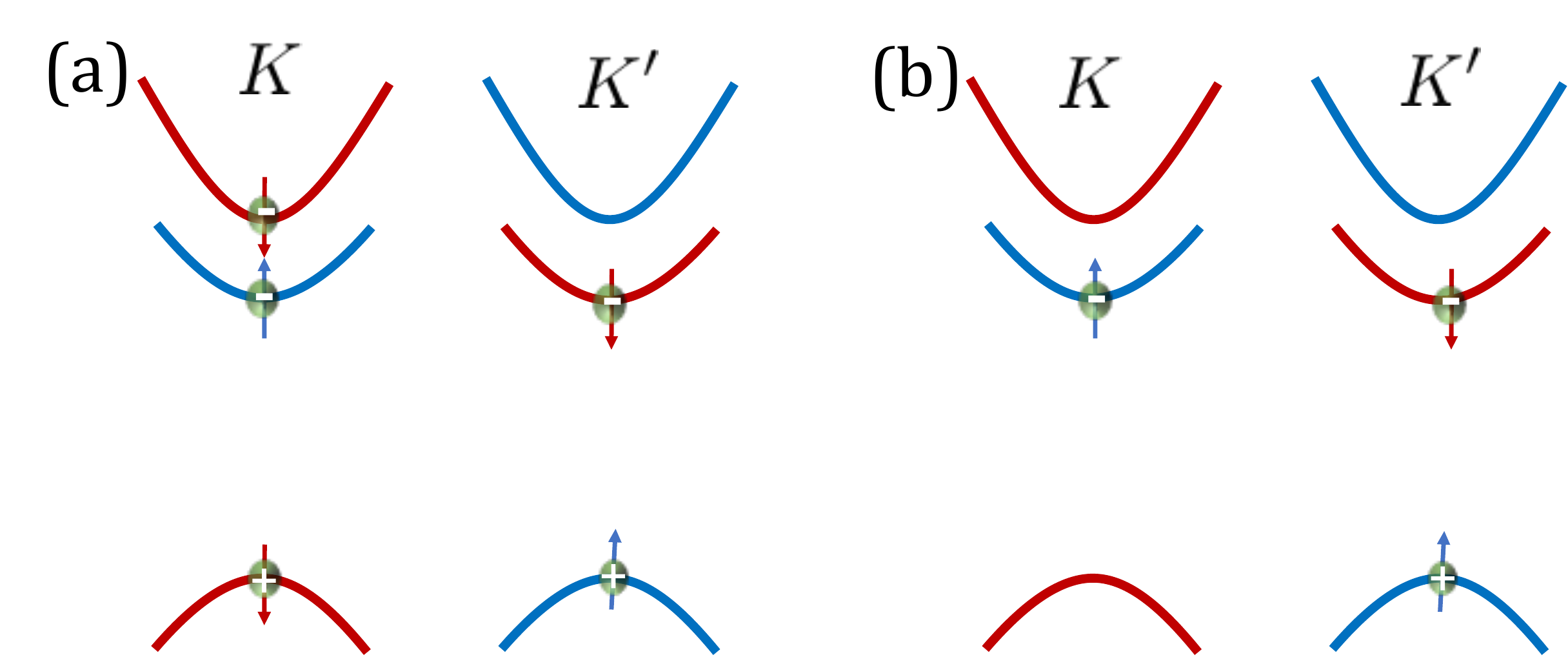}
 \caption{(a) The initial state in the recombination of a delocalized five-particle complex. This state  can be viewed as a charged biexciton or an exciton-trion. (b) The delocalized dark trion state after recombination.}\label{fig:delocal5}
\end{figure}

Given these findings, there are two possible interpretations to the physical origin of the peak $XX^-$ in the emission spectrum. The first one is that it stems from radiative recombination of a delocalized five-particle complex after which it turns to a dark trion, as shown in Fig.~\ref{fig:delocal5}. This interpretation was suggested in Refs.~[\onlinecite{Chen_NatComm18,Ye_NatComm18,Li_NatComm18,Barbone_NatComm18}], and it does not involve optical phonons. The second interpretation is that a localized electron mediates the phonon-assisted recombination of a neutral biexciton. Below, we first discuss the neutral exciton and biexciton states for which there a consensus, followed by analysis of the phonon-assisted recombination process. We then analyze the radiative recombination of the delocalized five-particle complex. 

\subsection{Radiative recombination of neutral biexcitons}\label{sec:exp_bi}
To evaluate the gained  energy when the electrons and holes interact, we consider the zero-energy reference level to be the top edge of the valence band. The energy gaps with the edges of the bottom and top valleys in the conduction band are denoted by $E_g$ and $E_g+\Delta$, respectively, where $\Delta$ is the spin-splitting energy in the conduction band (Fig.~\ref{fig:diagram}). 

Starting with the neutral exciton, the gained energy by the interaction is the electron-hole binding energy,  denoted by $E_{X^0}$ in the case of a bright exciton, and by $E_{X_D^0}$ in the case of a dark exciton. The photon energies due to the recombination of these excitons are
 \begin{eqnarray}
\hbar \omega_{X^0}    &=& E_g + \Delta - E_{X^0}  \ \,,  \label{Bind_Bexciton} \\
\hbar \omega_{X_D^0} &=&  E_g  - E_{X_D^0}.  \label{Bind_Dexciton}
\end{eqnarray}

Similarly, the gained energy when the four particles of a neutral biexciton interact is denoted by $E_{XX^0}$, 
 \begin{eqnarray}
 2E_g + \Delta -  E_{XX^0} &=& (E_g + \Delta - E_{X^0})   \nonumber \\ && \,\,\,+ \,\, (E_g - E_{X_D^0})   \, - \,\mathcal{E}_{XX^0}\,,
 \label{Bind_4Par1}
\end{eqnarray}
where $\mathcal{E}_{XX^0}$ is the binding energy of the biexciton, taken with respect to a system made of  bright and dark excitons that do not interact with each other. The photon energy due to the recombination of a neutral biexciton is 
 \begin{eqnarray}
\hbar \omega_{XX^0} &=& \hbar \omega_{X^0} - \mathcal{E}_{XX^0}  \ \,.  \label{eq:photon_Biexciton}
\end{eqnarray}

\subsection{Phonon-assisted recombination of a biexciton mediated by a localized electron}\label{sec:exp_5_phonon}
The phonon-assisted recombination has a strong signature in the emission spectrum when the following conditions are met.  The first condition is that the density of electrons in the ML is not much larger than the density of localization centers. When this condition is met, the attraction of the biexciton to these centers is not screened by itinerant electrons. The second condition is that the energy sum of the biexciton and localized electron (initial state) is close enough to the energy sum of the phonon and localized five-particle state (intermediate virtual state). We will show that this is the case in Sec.~\ref{sec:theory}, implying that the phonon-assisted recombination process is nearly a resonance process, which also explains the absence of emission from off-resonance higher phonon replica. In addition, we will show that the phonon-mediated interaction is strongest when the initial kinetic energy of the biexciton is relatively small and when the localized electrons are tightly bound (i.e., the effect is stronger when the impurities are closer to the ML because they induce tighter localization). 

Regardless of how strong the peak is in the emission spectrum, the energy conservation of this process requires that
 \begin{eqnarray}
&& \!\! (2E_g + \Delta -  E_{XX^0}) +  (E_g - E_{\ell}) =  \nonumber \\ && \hbar \omega_{XX^-}  +  E_{\text{ph}} 
  +  (E_g  - E_{X_D^0}) +  (E_g - E_{\ell}) .
 \label{Bind_5Par1}
\end{eqnarray}
The first line is the energy sum of the biexciton and localized electron in the initial state. The energy level of the localized electron is $E_g - E_{\ell}$,  where $E_{\ell}$ is the binding energy due to the interaction with the impurity.  The second line is the energy sum of the emitted photon ($\hbar \omega_{XX^-}$), phonon ($E_{\text{ph}}$), dark exciton, and localized electron in the final state.  Cancelling the localized electron terms on both sides and making use of Eq.~(\ref{Bind_4Par1}), we get
 \begin{eqnarray}
&& \!\!     (E_g + \Delta - E_{X^0}) + (E_g - E_{X_D^0})  - \mathcal{E}_{XX^0} =  \nonumber \\ && \hbar \omega_{XX^-} 
 +  E_{\text{ph}}  +  (E_g  - E_{X_D^0}) .
 \label{Bind_5Par2}
\end{eqnarray}
Cancelling the dark exciton terms on both sides and invoking Eqs.~(\ref{Bind_Bexciton}) and (\ref{eq:photon_Biexciton}), we arrive at what was expected all along,
 \begin{equation}
\Delta_{XX} \equiv  \hbar \omega_{XX^0} - \hbar \omega_{XX^-}  = E_{\text{ph}} .
 \label{Bind_5Par3}
 \end{equation}
Namely, the energy difference between the photon energies of the biexciton and five-particle peak is the phonon energy in the case that a localized electron mediates the phonon-assisted recombination.

At this point we mention that differences in charge densities, screening effects and fabrication conditions lead to slight changes in the nominal band-gap energies and binding energies.\cite{Meckbach_PRB18,Thygesen_2DMater17,Scharf_JPCM19} As was shown in Table \ref{tab:EPosition}, this fact is evident from the variations in the emitted photon energies that were observed in the four experiments.  However, the one consistent feature in Table~\ref{tab:EPosition} is the energy difference between the biexciton and five-particle  peaks: it is between 31 and 32 meV in all of the experiments. This finding reinforces the viability of the phonon-assisted recombination because the optical phonon energy is independent of the charge density, fabrication techniques, and dielectric screening.   

Additional strong support for the case of phonon-assisted recombination of biexcitons is that a similar behavior is seen for excitons. For example, Courtade \textit{et al.} have shown that the phonon-assisted recombination peak of excitons emerges 32~meV below $X^0$ in hBN-encapsulated ML-WSe$_2$.\cite{Courtade_PRB17} The phonon-assisted emission was evident only when the ML was neutral and it was suppressed when the gate-induced density of electrons was large (in which case the localization centers are screened out by delocalized electrons).  Furthermore, delocalized intravalley and intervalley trions emerge at elevated charge densities and their energies were 35 and 29~meV  below $X^0$, respectively. Their energy proximity to the  phonon-related peak is consistent with the resonance nature of the phonon-assisted recombination process.\cite{VanTuan_arXiv19} Finally, it is emphasized that the above analysis was based on crossing empirical results without invoking simulation results at any point. 

\subsection{Radiative recombination of delocalized five-particle complexes} \label{sec:Explan}

We now consider recombination from a delocalized five complex for which the valley and spin are good quantum numbers. As shown in Fig.~\ref{fig:delocal5}(a), this complex comprises two holes and three electrons, and it can be viewed as a delocalized charged biexciton or exciton-trion complex. After recombination of its bright exciton component, the complex turns into a dark trion as shown in Fig.~\ref{fig:delocal5}(b). Denoting the gained energy from the interaction of the five particles in the initial state by $E_{XX^-}$, and from the interaction of the three particles in the dark-trion final state by $E_{X_D^-}$, we get that 
 \begin{eqnarray}
3E_g + \Delta -  E_{XX^-} &=& (E_g + \Delta - E_{X^0}) \nonumber \\ &&\,\, +  \,\,(2E_g -  E_{X_D^-})   \,\,  -  \,\,\mathcal{E}_{XX^-} \,\,\,\,.
 \label{Bind_deloc_5Par1}
\end{eqnarray}
$\mathcal{E}_{XX^-}$ is the binding energy of the five-particle complex, taken with respect to a delocalized system made of  a bright exciton and dark trion that do not interact with each other. Making use of Eq.~(\ref{Bind_Bexciton}), the photon energy due to recombination of the bright exciton component in the five-particle complex is
\begin{eqnarray}
\hbar \omega_{XX^-}    &=& \hbar \omega_{X^0}     \,\, - \,\, \mathcal{E}_{XX^-}  \ \,.  \label{photon_5deloc}
\end{eqnarray}
In the final step, we make use of Eqs.~(\ref{eq:photon_Biexciton}) and (\ref{photon_5deloc}) to evaluate the difference between Eqs.~(\ref{Bind_4Par1}) and (\ref{Bind_deloc_5Par1}). We get that 
 \begin{eqnarray}
E_{XX^-} - E_{XX^0}&=&   \mathcal{E}_{X_D^-}  \,+\,    \Delta_{XX}  \,\,\,\,.
 \label{eq:final_deloc5}
\end{eqnarray}
$\mathcal{E}_{X_D^-} = E_{X_D^-} - E_{X_D^0}$ is the binding energy of the dark trion, taken with respect to a noninteracting free electron and dark exciton. $\Delta_{XX} = \hbar \omega_{XX^0}  - \hbar \omega_{XX^-}$ is the energy difference between the photon energies of the biexciton and five-particle peaks. The only way to confirm whether the recombination of the delocalized five-particle complex is a possible explanation for the peak $XX^-$ is through simulations that can check the correctness of Eq.~(\ref{eq:final_deloc5}).

\begin{table} [t!]
\renewcommand{\arraystretch}{1.25}
\tabcolsep=0.10 cm
\caption{\label{tab:E_deloc} Calculated gained and binding energies of the bright exciton, dark exciton, dark trion, neutral and charged biexcitons in hBN-encapsulated ML-WSe$_2$. The expressions to calculate the binding energies are also shown. All complexes are delocalized.}
\begin{tabular}{|c|c||l|c|}
\hline\hline
Energy 			& meV 	             		&    Binding energy     									&  meV  \\ \hline
$E_{X^0}$ 		& 178 			&   $\mathcal{E}_{X^0}     = E_{X^0} $							& 178		\\  
$E_{X_D^0}$ 		& 195.2			&   $\mathcal{E}_{X_D^0} = E_{X_D^0} $		 					&  195.2 		\\ 
$E_{X_D^-}$  		& 220.8 			&   $\mathcal{E}_{X_D^-}  = E_{X_D^-} - E_{X_D^0}$				&  25.6		\\  
$E_{XX^0}$  		& 390.6 			&   $\mathcal{E}_{XX^0}  = E_{XX^0} -  E_{X^0} - E_{X_D^0}$		   	& 17.4   	         \\ 
$E_{XX^-}$		& 441.5  			&   $\mathcal{E}_{XX^-}  = E_{XX^-}   -  E_{X^0} - E_{X_D^-}$		         &  42.7              \\
\hline \hline
\end{tabular}
\vspace{-2mm}
\end{table}

Table~\ref{tab:E_deloc} lists the simulation results, which we got by employing the stochastic variational method (SVM).\cite{Suzuki_Varga_Book98,Varga_CPC97,Varga_PRC95,Varga_CPC08,Mitroy_RMP13}  Ref.~[\onlinecite{VanTuan_PRB18}] as well as Appendix~\ref{app:SVM} include technical details on the computation methods, and Appendix~\ref{app:parameters} lists the parameter values we use in the simulations. Appendix~\ref{app:benchmark_delocalized} includes a discussion in which the calculated values in Table~\ref{tab:E_deloc} are benchmarked against experimental results. Substituting the calculated values for $E_{XX^-}$, $E_{XX^0}$ and $\mathcal{E}_{X_D^-}$ into Eq.~(\ref{eq:final_deloc5}), and assuming that $\Delta_{XX} = 32$~meV (Table~\ref{tab:EPosition}), we get a mismatch of $\sim$6~meV.

All in all, we have discussed two possible interpretations for the peak $XX^-$ seen in experiments.\cite{Chen_NatComm18,Ye_NatComm18,Li_NatComm18,Barbone_NatComm18}  The explanation that this peak stems from recombination of delocalized five-particle complexes can only be verified through simulations of the involved states, and the energy conservation of this process resulted in a slight mismatch that one can attribute to the choice of parameters. On the other hand, the explanation that this peak stems from localized electrons that mediate the phonon-assisted recombination of neutral biexcitons can be verified without any calculation. The one requirement is that the phonon energy is the energy difference between the photon energies of the biexciton and five-particle peaks. The energy difference between these photons is $\sim$32~meV in hBN/ML-WSe$_2$/hBN,\cite{Chen_NatComm18,Ye_NatComm18,Li_NatComm18,Barbone_NatComm18} which is consistent with the phonon energy found in other experiments.\cite{Courtade_PRB17}  While we favor the phonon-assisted recombination process, more experiments are needed to verify which of the two proposed processes is correct (e.g., by using $\mu$-PL to probe whether the emission of the five-particle peak is localized in nature compared with the spatially uniform emission of the neutral exciton).

 \section{Microscopic details of the phonon-assisted recombination process} \label{sec:theory}
 
We have shown that the spectral position of the five-particle peak in experiments is what one should expect from the phonon-assisted recombination process. In this section we provide microscopic details of the interaction between the biexciton and localized electron through which one can evaluate the relative amplitude of the peak in the emission spectrum. 

 \subsection{Capture process}\label{sec:CaptureProcess}
We begin by analyzing the matrix element of the capture process, focusing on the Fr\"{o}hlich interaction with long-wavelength LO phonons ($E_2'$) because of its dominant effect in polar materials.\cite{Kaasbjerg_PRB12,Sohier_PRB16} The matrix element involves the interaction between each of the particles with the phonon-induced macroscopic field. When the system comprises three particles (exciton and localized electron), the matrix element reads\cite{VanTuan_arXiv19}  
\begin{eqnarray}
\!\!\!\! \!\!\mathcal{M}_{3,\lambda}(\mathbf{K},\mathbf{q}) \!=\!  \langle \Psi_{X^-}  | \, \Sigma_j  D_{j,\lambda}(\mathbf{q})e^{i\mathbf{q}\mathbf{r}_j}  | \Psi_{X^0}(\mathbf{K})\Psi_{\ell}   \rangle,\,\,\,\,\, \label{eq:M3}
\end{eqnarray}
Similarly, the matrix element for a five-particle system (biexciton and localized electron) reads
\begin{eqnarray}
\!\!\!\! \!\!\mathcal{M}_{5,\lambda}(\mathbf{K},\mathbf{q}) \!=\!  \langle \Psi_{XX^-}  | \, \Sigma_j  D_{j,\lambda}(\mathbf{q})e^{i\mathbf{q}\mathbf{r}_j}  | \Psi_{XX^0}(\mathbf{K})\Psi_{\ell}   \rangle.\,\,\,\,\, \label{eq:M5}
\end{eqnarray}
In both cases, $\lambda$ and $\mathbf{q}$ are the phonon mode and wavevector, respectively. $\mathbf{K}$ is the wavevector of the exciton or biexciton in the initial state. The wavefunctions of the initial state are the localized electron ($\Psi_{\ell}$) and delocalized exciton or biexciton ($\Psi_{X^0}$ or $\Psi_{XX^0}$). The wavefunction of the state that ensues the capture process is a localized trion or charged-biexciton  ($\Psi_{X^-}$ or $\Psi_{XX^-}$). The sum runs over the particles of the system whose positions are denoted by $\mathbf{r}_j$. $D_{j,E_2'}(q)$ is the coupling between the phonon-induced macroscopic  electric field and the $j^{\text{th}}$ particle.  Appendix~\ref{app:SVM} provides details of the wavefunctions, Fr\"{o}hlich interaction, and the calculation technique used to evaluate the matrix element. Appendix~\ref{app:parameters} is a summary of the parameters we have used in the simulations. 

One can recognize the importance of having an uneven number of electrons and holes in the system from the fact that the electron-phonon and hole-phonon Fr\"{o}hlich couplings only differ in their sign. As a result, the sum $\Sigma_j  D_{j,\lambda}(\mathbf{q}) e^{i\mathbf{q}\mathbf{r}_j}$ in Eqs.~(\ref{eq:M3}) and (\ref{eq:M5}) vanishes when dealing with neutral excitons or biexcitons in the long-wavelength limit,  $q \rightarrow 0$.  In fact, the matrix element vanishes identically when the particles of the neutral complex have the same mass. By adding the localized electron into the mix, on the other hand, the system is no longer neutral and the sum becomes finite in the long-wavelength limit. 

\begin{figure}
 \centering
\includegraphics[width=8.5 cm]{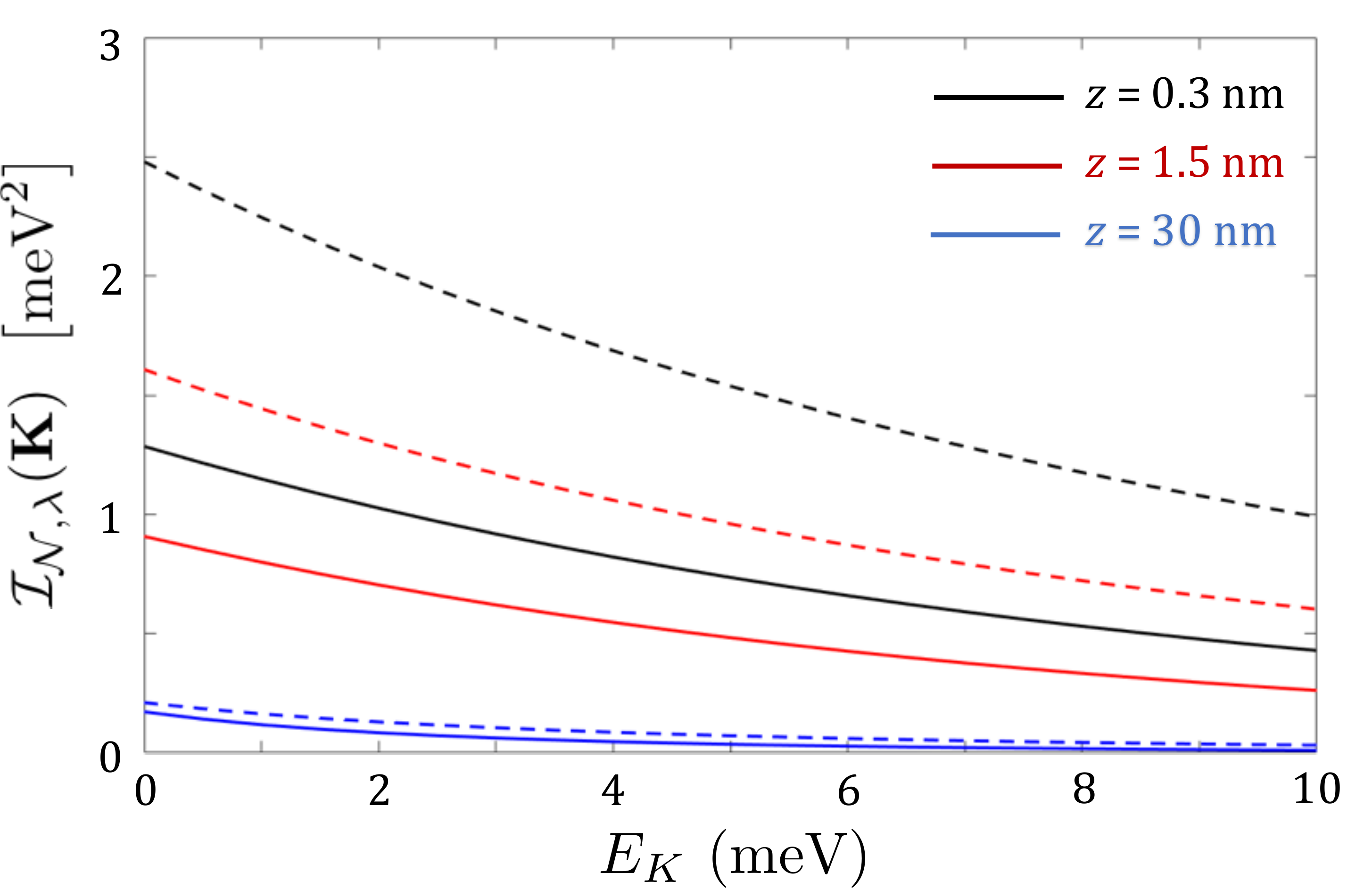}
 \caption{The integrated square matrix elements for phonon-assisted recombination in hBN/ML-WSe$_2$/hBN as a function of kinetic energy of the biexciton (dashed lines) and exciton (solid lines). Results are shown for three different distances of the charged impurities from the ML. The impurity density is $n_d= 2\times10^{11}$~cm$^{-2}$.  }\label{fig:IK}
\end{figure}

Next, we sum over contributions from all possible phonon wavevectors and factor the integrated matrix element by the number of localized electrons. The latter is denoted by $n_dA/2$, where $n_d$ is the density of localized electrons, $A$ is the area of the sample, and the factor 1/2 is due to the valley-mixing effect (only half of the excitons or biexcitons can interact with the localized electron). The integrated square matrix elements are defined as,
\begin{eqnarray}
\mathcal{I}_{\mathcal{N},\lambda}(\mathbf{K}) &\equiv&    \frac{n_dA}{2} \sum_{ \mathbf{q}}   \left| \mathcal{M}_{\mathcal{N},\lambda}(\mathbf{K},\mathbf{q}) \right |^2    \,. \label{eq:Ik}
\end{eqnarray}  
$\mathcal{N}=3~(5)$ when the localized electrons interact with excitons (biexcitons). Figure~\ref{fig:IK} shows simulation results of $\mathcal{I}_{3,\lambda}(\mathbf{K})$ in solid lines and $\mathcal{I}_{5,\lambda}(\mathbf{K})$ in dashed lines for hBN/ML-WSe$_2$/hBN  as a function of the exciton (biexciton) kinetic energy, $E_K = \hbar^2K^2/2M$. $M$ is the sum of effective masses of the particles that compose the exciton (biexciton). Results are shown for three different distances of the remote charged impurities from the ML.

Figure~\ref{fig:IK} reveals two important trends. The first one is that the capture process is stronger for nearby impurities as evident from the amplitudes of the three dashed or solid lines. This result reinforces the important role of localization since nearby impurities induce tighter localization. The latter is needed because the capture process is  optimal when the localization length of the electron is comparable to the typical  distance between the particles in the localized trion (or charged biexciton) state. The second trend seen in Fig.~\ref{fig:IK} is  the monotonous decay of the matrix element when the kinetic energy of the exciton (or biexciton) increases. Namely, the capture is less effective at elevated temperatures or immediately after photoexcitation because it is harder to capture fast/hot excitons. Nonetheless, Fig.~\ref{fig:IK} shows that the capture is still effective when the kinetic energy is a few meV, which is orders of magnitude larger than the kinetic energy of excitons in the minuscule light cone (a few $\mu$eV). Therefore, the number of excitons that  can take part in the phonon-assisted process far exceeds that of excitons that can spontaneously radiate.   

 \subsection{Recombination Rate}\label{sec:RR}

Having the integrated matrix element of the capture process, the phonon-assisted recombination rate follows from\cite{VanTuan_arXiv19}
\begin{eqnarray}
\!\!\!\!\!\! \frac{1}{\tau_{\mathcal{N},\lambda}(\mathbf{K})} &=&    \frac{ 1 \pm p_{\ell} } {\tau_{\mathcal{N},\ell}}  \cdot \frac{ \mathcal{I}_{\mathcal{N},\lambda}(\mathbf{K})  }{ (\Delta E_{\mathcal{N}} + E_{\lambda} - E_{\mathbf{K}})^2 +\Gamma^2 }  \,, \label{eq:tau_p}
\end{eqnarray}  
$E_{\lambda}$ is the phonon energy, $\Gamma$ is a broadening parameter, and $\tau_{\mathcal{N},\ell}$ is the radiative decay time of a localized $\mathcal{N}$-particle complex. The term $1 \pm p_{\ell}$ denotes the change in the recombination rate when the localized electron system becomes spin-polarized ($p_{\ell} \neq 0$). We will elaborate on this case in the next part when we study the dependence of the phonon-assisted recombination process on magnetic field. Finally, $\Delta E_{\mathcal{N}}$ is the difference between the energies of the few-body complexes at the initial and intermediate states,
\begin{eqnarray}
\Delta E_{\mathcal{N}} =  \left|\, E_{\mathcal{N}-1} + E_{\ell} \, \right|  - |E_{\mathcal{N},\ell}| \,. \label{eq:dEN}
\end{eqnarray}  
The intermediate state is a localized trion when $\mathcal{N}=3$  ($E_{\mathcal{N},\ell} = E_{X^-,\ell}$) or a localized charged-biexciton when $\mathcal{N}=5$  ($E_{\mathcal{N},\ell} = E_{XX^-,\ell}$). The initial-state energies are that of the localized electron,  $E_{\ell}$, and delocalized $(\mathcal{N}-1)$-particle complex. The latter is an exciton when $\mathcal{N}=3$  ($E_{\mathcal{N}-1} = E_{X^0}$) or a biexciton when $\mathcal{N}=5$  ($E_{\mathcal{N}-1} = E_{XX^0}$).

Table~\ref{tab:E_L} shows the calculated gained energies of the localized complexes that are involved in the recombination process. Using $E_{\lambda}=32$~meV, and the calculated energies of the delocalized complexes, $E_{X^0}=178$~meV and $E_{XX^0}=390.6$~meV from Table~\ref{tab:E_deloc}, we can extract the values of $\Delta E_{3}+E_{\lambda}$ and $\Delta E_{5}+E_{\lambda}$. Their relatively small values, as listed in Table~\ref{tab:E_L}, demonstrate the resonance nature of the phonon-assisted recombination process.  

\begin{table} [t!]
\renewcommand{\arraystretch}{1.25}
\tabcolsep=0.10 cm
\caption{\label{tab:E_L} Calculated energies in meV of the localized electron, trion, and charged biexcitons in ML-WSe$_2$ encapsulated in hBN. The results are shown for different distances of the impurities from the mid-plane of the ML. Also shows are the values of $\Delta E_{3}+E_{\lambda}$ and $\Delta E_{5}+E_{\lambda}$ (see text).}
\begin{tabular}{|c||c|c|c||c|c|}
\hline\hline
$z$ (nm) 		& 	$E_{XX^-,\ell}$ 		&    	$E_{X^-,\ell}$    	&	$E_{\ell}$ 	         &    	$\Delta E_{3}+E_{\lambda}$ 		&  	$\Delta E_{5}+E_{\lambda}$  		\\ \hline
0.3			& 	542.8			&   	315.1			&	110.8		&      5.7				&      -9.4			\\  
1.5 			& 	512.0			&   	284.5	 		&	75.8			&      1.3 		        		&      -13.6			\\ 
30  			&	446.4 			&   	220.2			&      10.4			&      0.2 				&      -13.4			\\  
$\infty$  		&	441.5 			&   	213.0	   	         &     0  	        		&      -3.0                 		&      -18.9  			\\ 
\hline \hline
\end{tabular}
\vspace{-2mm}
\end{table}

To evaluate the amplitude of the phonon-assisted recombination peak in the emission spectrum, one has to know the radiative rates of delocalized and localized complexes (without phonons). In addition, 
non-radiative recombination rates may be different for localized and delocalized complexes. Instead of assuming the values of many fitting parameters, we will quantify the change  in amplitude of the phonon-assisted recombination process in the presence of a magnetic field. The calculated behavior can then be directly benchmarked against the trends seen in experiments.

\begin{figure*}
 \centering
  \begin{minipage}{.48 \textwidth}
\includegraphics[width=9cm]{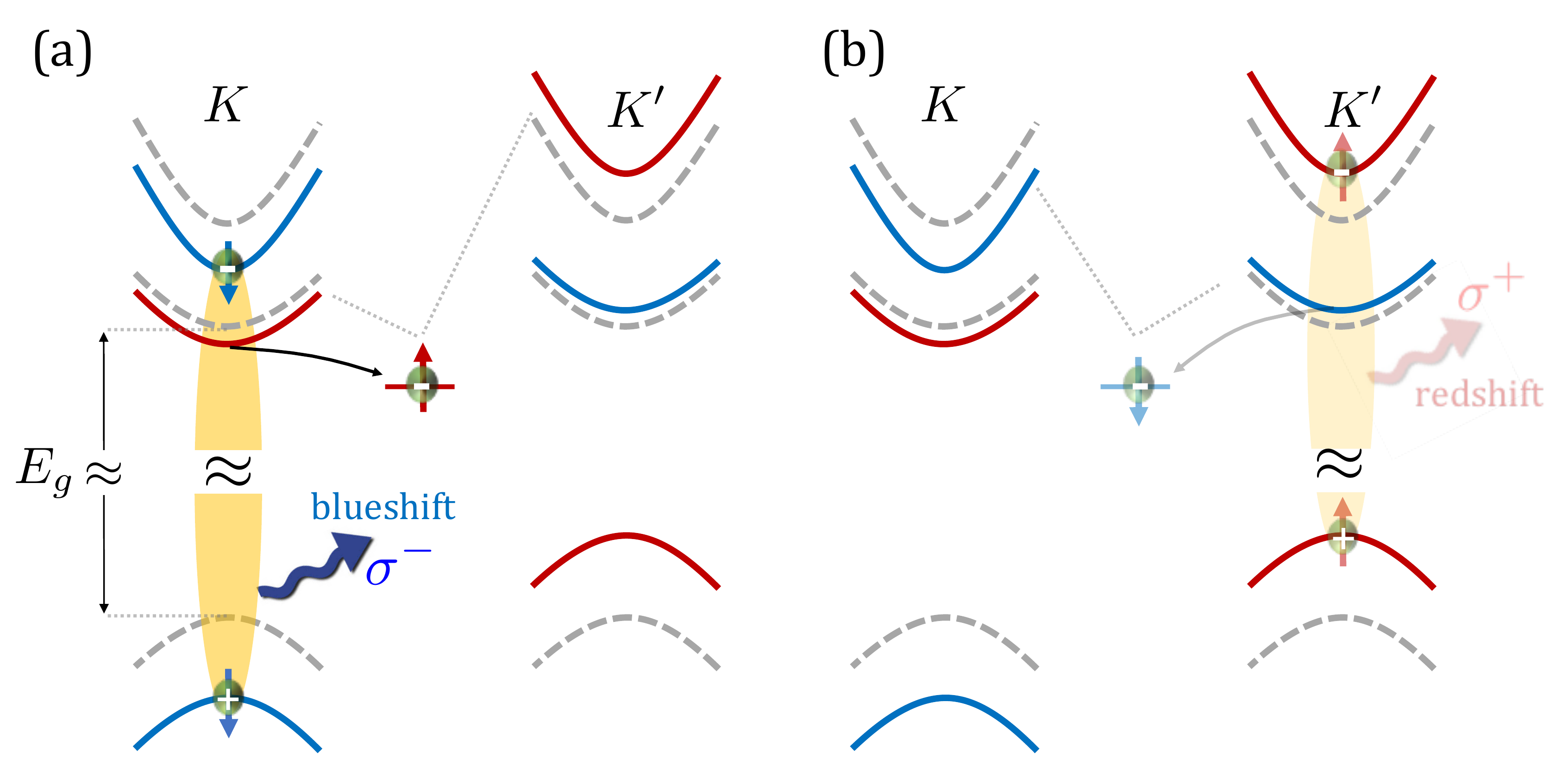}
 \end{minipage}
  \hfill
    \begin{minipage}{.48 \textwidth}
\includegraphics[width=9cm]{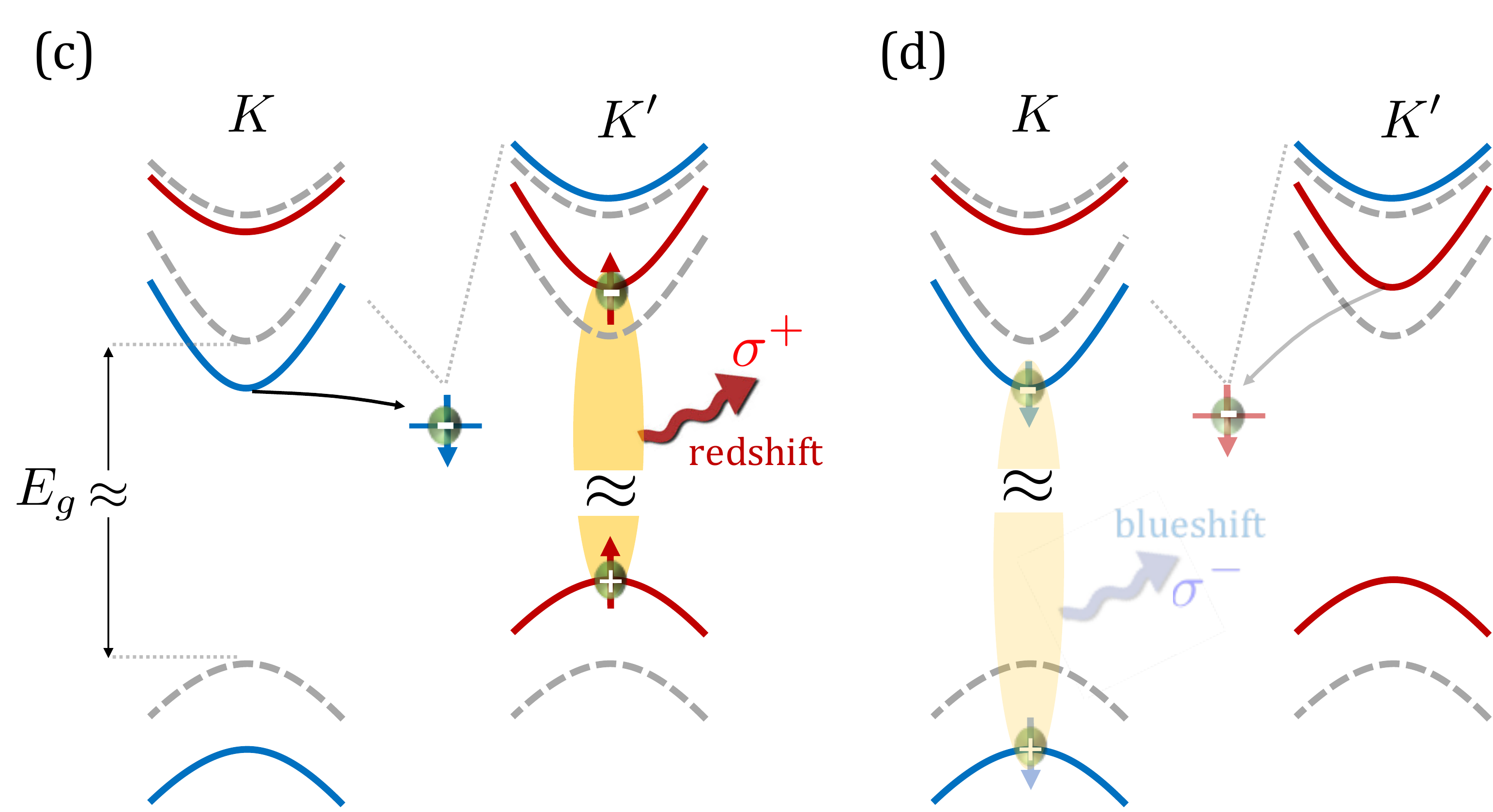}
 \end{minipage}
  \hfill
\caption{The initial states of the phonon-assisted recombination process in the presence of a magnetic field. The case of ML-WSe$_2$ is shown in (a) and (b), while ML-MoSe$_2$ in (c) and (d). The dashed-grey lines are the valleys energy levels at zero field, and the solid color lines show their displaced positions after applying the magnetic field.  The circular polarization of the emitted photons in the $K$- and $K'$-point valleys are chosen as $\sigma^-$ and $\sigma^+$, respectively. Localized electrons assume their spin polarization from that of free electrons in the bottom valleys of the conduction band, as indicated by the arrows between these valleys and the localized electrons. The coupling between excitons and the majority of localized electrons is shown in (a) for ML-WSe$_2$ and in (c) for ML-MoSe$_2$. The respective couplings between excitons and the minority of localized electrons are shown in (b) and (d). The PL associated with the majority (minority) population intensifies (fades out) when the amplitude of the magnetic field increases.}\label{fig:B_diagram_trion}
\end{figure*}
 \subsection{The magnetic-field dependence of the phonon-assisted recombination process}
 
Application of an out-of-plane magnetic field in ML-TMDs splits the peaks in the emission spectrum to blueshifting or redshifting branches due to the  Zeeman-induced energy shift at the edges of the conduction- and valence-band valleys. A similar effect occurs when the ML is placed in proximity to magnetic materials.\cite{Zutic_PM19} Denoting the total density of localized electrons by $n_d$, the majority and minority spin densities are $(1 + p_{\ell} )n_d/2$ and $(1 - p_{\ell} )n_d/2$, respectively, where $p_\ell$ is the polarization magnitude ($0< p_{\ell} < 1$). 

The spin polarization of localized electrons in the presence of a magnetic field is established when they are captured or through exchange interactions with spin-polarized delocalized electrons.\cite{Qing_PRB15} In devices where the gate voltage is tuned to the neutral regime, delocalized electrons and holes are introduced by photo-doping (i.e., electrons and holes from dissociated excitons). Here, we assume that exchange interactions between localized and delocalized electrons  are more effective in polarizing the localized electrons compared with the relatively slow spin-lattice processes that flip the spin of an isolated localized electron.\cite{Qing_PRB15} The ratio between the majority and minority spin densities of localized electrons is then approaching the respective ratio of the delocalized electron system. Assuming a small density of delocalized electrons such that $E_F < k_BT$, and denoting the effective $g$-factor in the bottom of the conduction band by $g_{\text{cb}}$, we can then write that
\begin{equation}
\frac{1+p_{\ell}}{1-p_{\ell}} =  \exp\left( \frac{\left| g_{\text{cb}}\mu_B B \right| }{k_BT} \right).
\label{Eq:Polar1}
\end{equation}
This result is appealing because one can use the magnetic-field dependent PL to extract the  $g$-factor of electrons (or holes) and not only of excitons or trions. We explain this point in more detail below.  

As we have seen in Sec.~\ref{sec:localization}, a localized electron can be in contact with an exciton whose electron has opposite spin due to the Pauli exclusion principle and localization-induced valley mixing. This behavior is depicted in Fig.~\ref{fig:B_diagram_trion}, where the magnetic-field induced energy shifts of the valleys follow the $g$-factor analysis in Ref.~[\onlinecite{Koperski_2DMater19}].  Figures~\ref{fig:B_diagram_trion}(a) and (b) correspond to the interaction between excitons and localized electrons from the majority and minority spin populations in ML-WSe$_2$, respectively. A similar behavior is expected in ML-WS$_2$. Figure~\ref{fig:B_diagram_trion}(a) shows that the blueshifting excitons are coupled to the majority population of localized electrons (i.e., localized electrons that accept their spin polarization from the bottommost valley in the conduction band whose energy is lowered by the magnetic field). Conversely, Fig.~\ref{fig:B_diagram_trion}(b) shows that the redshifting excitons are coupled to the minority population. The cases of ML-MoSe$_2$, shown in Figs.~\ref{fig:B_diagram_trion}(c) and (d) are similar, only  that now the redshifting and blueshifting  branches are switched due to the reversed order of the  top and bottom valleys in the conduction band. 

By continuing to increase the magnetic field, delocalized electrons populate only the bottommost valley in the conduction band, and the localized electrons eventually become fully polarized ($p_{\ell} \rightarrow 1$ when $|g_{\text{cb}}\mu_B B | \gg k_BT$). The net effect is that the phonon-assisted recombination of excitons in tungsten-based MLs is enhanced for  the blueshifting branch and suppressed for the redshifting one, as shown in Fig.~\ref{fig:B_diagram_trion}(a) and (b). This behavior explains a recent  low-temperature PL experiment of ML-WS$_2$ supported on SiO$_2$.\cite{Plechinger_NanoLett16} The phonon-assisted peak, denoted by $X_U$ in Fig.~1 of Ref.~[\onlinecite{Plechinger_NanoLett16}], was evidently amplified in the blueshifting branch when the magnetic field was ramped-up from zero to 30~Tesla (and the phonon-assisted recombination process was suppressed in the redshifting branch). 

The same behavior is expected for biexcitons in tungsten-based MLs because their dark exciton components are left behind after the recombination process. In other words, while one can simply add the dark exciton components in the $K$'-point valley of Fig.~\ref{fig:B_diagram_trion}(a) and the $K$-point valley of Fig.~\ref{fig:B_diagram_trion}(b), the energy of the emitted photon is governed by recombination of the bright exciton component.  As before, the net result is that the phonon-assisted recombination of biexcitons in tungsten-based MLs is enhanced for  the blueshifting branch and suppressed for the redshifting one. This physical picture is consistent with the findings  of Refs.~[\onlinecite{Chen_NatComm18,Ye_NatComm18,Li_NatComm18,Barbone_NatComm18}]. Finally, we mention that by reversing the sign of the magnetic field, the only expected change is opposite circular polarization in the PL. The blueshift and redshift analysis remains the same. Namely, the blueshifting (redshifting) branch of the phonon-assisted optical transition when mediated by localized electrons is amplified (suppressed) in tungsten-based MLs regardless of the sign of the out-of-plane magnetic field. 

An additional important point is that the above magnetic-field dependence remains the same for photon emission from localized trions (i.e., without phonons). All of these processes involved localization, and one can extract the $g$-factor of electrons in the bottom of the conduction band by fitting Eq.~(\ref{Eq:Polar1}) to the measured intensity ratio between the amplified blueshifting and suppressed redshifting peaks in tungsten-based MLs.  The case of ML-MoSe$_2$ is subtle because the energy levels of positive and negative trions are much closer to each other.  If negative complexes can be resolved in high-quality ML-MoSe$_2$ devices, then Eq.~(\ref{Eq:Polar1}) should be matched to the intensity ratio between the amplified redshifting and suppressed blueshifting branches of localized trions.  If positive complexes are resolved, then one can also extract the $g$-factor of holes in the top valley of the valence band [$g_{\text{cb}}\,\rightarrow\,g_{\text{vt}}$ in Eq.~(\ref{Eq:Polar1})]. Both in tungsten and molybdenum based compounds, the blueshifting peak of positive localized complexes should be amplified if we assume that the spin polarization of localized holes matches that of delocalized holes from the topmost valleys. This behavior can be inferred from the energy shifts in Fig.~\ref{fig:B_diagram_trion}, where one should now require that a localized hole is in contact with an exciton whose hole has opposite spin.  

\begin{figure}
 \centering
\includegraphics[width=8.5 cm]{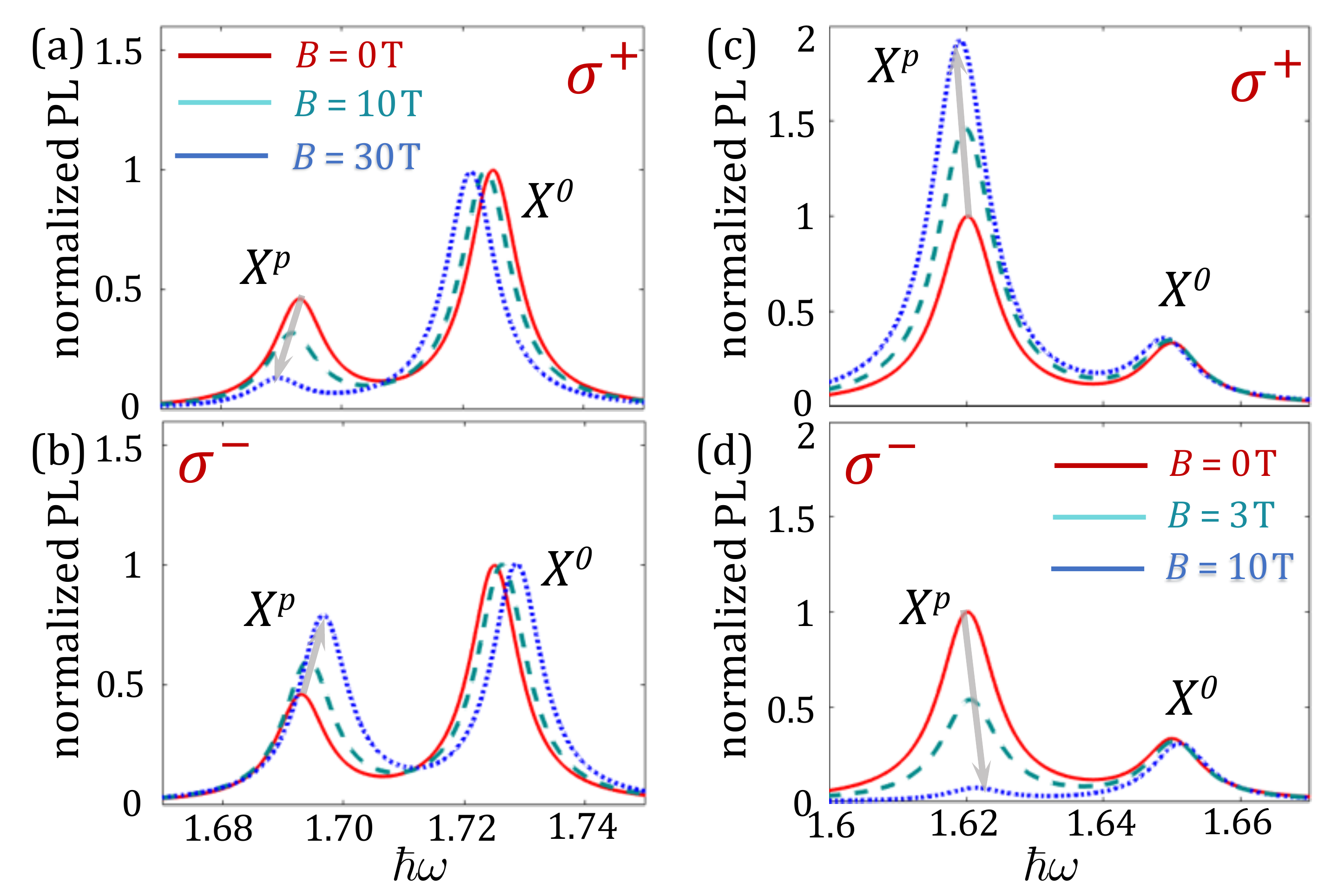}
 \caption{(a) and (b) The calculated PL of hBN/ML-WSe$_2$/hBN at $T$=10~K. The emission spectrum is shown separately for the $K$-point valley ($\sigma^-$) and $K$'-point valley ($\sigma^+$) at three different magnetic fields (0, 10 and 30~T).  (c) and (d) The respective calculations for hBN/ML-MoSe$_2$/hBN with 0, 3 and 10~T. The black arrows trace the phonon-related peaks and they serve as a guide to the eye.} \label{fig:PL_B} 
\end{figure}

Next, we quantify the magnetic field dependence of the phonon-assisted recombination process of excitons. Figure~\ref{fig:PL_B} shows the calculated PL of hBN/ML-WSe$_2$/hBN and hBN/ML-MoSe$_2$/hBN. The optical transitions in the $K$- and $K$'-point valleys are denoted by the emitted circularly polarized light $\sigma^-$ and $\sigma^+$, respectively. The two peaks in the emission spectrum are the neutral exciton, $X^0$, and its phonon-assisted recombination, $X^p$. The intensity of the former is commensurate with\cite{Wang_PRB16,Robert_PRB16}
\begin{eqnarray}
I_0 \sim   \left[ \frac{ (\hbar\omega_{X^0})^2 }{2 k_BT Mc^2 } \right] \frac{1}{\tau_X} \,, \label{eq:tau_o}
\end{eqnarray}
where $M$ is the  exciton mass, $c$ is the speed of light in vacuum and $k_BT$ is the thermal energy. $\tau_X$ is the radiative decay time of delocalized excitons in the light cone. The prefactor in the bracket takes into account the probability to find an exciton in the light cone. We assume that the recombination lifetime of excitons is faster than the time it takes for excitons from the $K$-point valley to reach mutual equilibrium with excitons from the $K'$-point valley. As a result, the only effect of the magnetic field is through the energy splitting of $\hbar\omega_{X^0}$, which is modeled in this example through the empirically known $g$-factor of bright excitons  ($|g_b| \approx 4.2$).\cite{Koperski_2DMater19}

The intensity of the phonon-assisted recombination of excitons is commensurate with
\begin{eqnarray}
 I_{\text{p}} \sim   \frac{\sum_{\mathbf{K}} f_X(\mathbf{K})/\tau_{\mathcal{N},\lambda}(\mathbf{K})}{\sum_{\mathbf{K}} f_X(\mathbf{K})} , \label{eq:avg}
\end{eqnarray}
where $f_X(\mathbf{K})$ is the distribution function of delocalized excitons with kinetic energy $E_K$. The rate of the phonon-assisted recombination process, $1/\tau_{\mathcal{N},\lambda}$  was evaluated from Eq.~(\ref{eq:tau_p}). The polarization of localized electrons follows from Eq.~(\ref{Eq:Polar1}), and we have modeled the behavior by assuming that $g_{\text{cb}}=1$ for ML-WSe$_2$ and $g_{\text{cb}}=5$ for ML-MoSe$_2$. These values are still not known empirically in the regime that the delocalized electron density is small.  We have assumed that $g_{\text{cb}}$ is larger in ML-MoSe$_2$ than in ML-WSe$_2$  because of the constructive versus destructive sum from valley-orbit and spin-orbit contributions. Other parameters used in Fig.~\ref{fig:PL_B} are $\tau_{3,\ell} = 10 \tau_X$ and we have assumed that impurities are $z=0.3$~nm away from the ML and their density is $n_d=2 \times 10^{11}$~cm$^{-2}$. Importantly, the ratio between the amplified and suppressed peaks in magnetic-field dependent PL experiments is influenced only by the ratio, $(1+p_{\ell})/(1-p_{\ell})$. If our assumption that this ratio is similar to that of delocalized electrons is correct, then one can use Eq.~(\ref{Eq:Polar1}) to extract the real $g$-factor values in the bottom valleys of the conduction bands.

Finally, we mention that the effect of the magnetic field on the matrix elements in Eq.~(\ref{eq:M3})-(\ref{eq:M5}) is relatively weak. Other than a  small diamagnetic shift, the exciton's ground state is hardly affected by the field because of its small size and neutrality.\cite{Stier_NanoLett16}  We have verified this result by adding the magnetic-field effect to the SVM model. We have also verified that the magnetic field hardly affects the localized electron or trion because of their relatively large binding energy to the localization center. The calculated effect is that the localization radius decreases by only $\sim$10\% when the magnetic field is  as high as 30 Tesla. The resulting change in the amplitude of the matrix element affects the recombination rate only marginally compared with the change induced by the polarization of the localized-electron system. 

\subsection{Coupling between homopolar and LO phonon modes} \label{sec:zolo}

Another phenomenon that we point to in this work is an intriguing coupling between the LO and homopolar phonons in ML-TMDs (The LO mode is also refereed to by $E_2'$ and the homopolar by $A_1'$ or ZO). Both Raman and PL spectra show that the dominant phonon is the mode with lower energy: LO in WS$_2$ and MoS$_2$, and the homopolar in WSe$_2$ and MoSe$_2$.\cite{VanTuan_arXiv19} The question we are raising is what causes this difference. Namely, if the Fr\"{o}hlich interaction is the strongest, then why not seeing a dominant signature of the LO mode in all ML-TMDs. We provide a possible explanation below. 

\begin{figure}
 \centering
\includegraphics[width=8.5 cm]{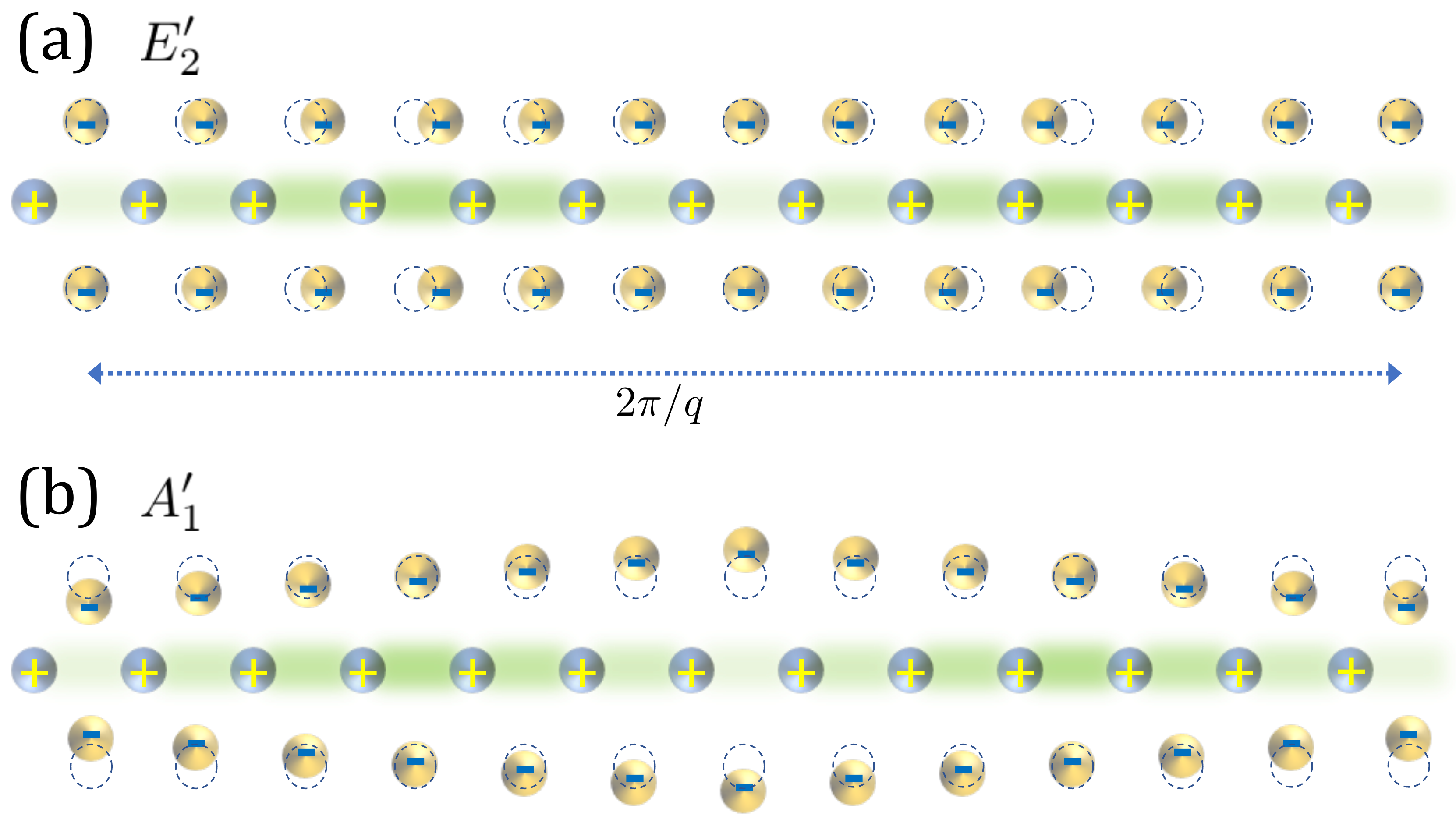}
 \caption{A cartoon of the atomic displacements of (a) the longitudinal-optical phonon mode, $E_2'$, and (b) the homopolar mode, $A_1'$. The dashed empty circles denote the equilibrium positions of the chalcogen atoms. The `green' clouds in the mid-plane represents the macroscopic in-plane electric field that is induced by these atomic displacements.}\label{fig:zolo}
\end{figure}

The LO mode is governed by the atomic displacement shown in Fig.~\ref{fig:zolo}(a). The in-phase motion of the top and bottom chalcogen atoms is out-of-phase with respect to the transition-metal atom.  Similar to other polar crystals, a macroscopic polarization is developed when the out-of-phase motion of dissimilar atoms in the unit-cell is directed along the wavevector  of the long-wavelength optical-phonon ($\mathbf{q} \parallel \mathbf{P}$).\cite{Cardona_Book} So far, the vast majority of studies and textbooks consider the LO mode as the origin for a long-range (Fr\"{o}hlich) interaction between optical phonons and itinerant charged particles in crystals.\cite{Frohlich_AP54,Devreese_RPP09}

Contrary to LO phonons, transverse optical (TO) phonons can couple to electrons or holes only through short-range interactions (when symmetry allows it). The cause for this interaction is essentially the volume change of the unit cell. Breathing phonon modes in crystals with more than two atoms per unit cell are an exception to this rule.\cite{Peelaers_PRB15} In ML-TMDs, this phonon is the homopolar mode governed by an out-of-phase motion of the chalcogen atoms in the out-of-plane direction, as shown in Fig.~\ref{fig:zolo}(b). To date, the coupling between electrons or holes due to the thickness fluctuations induced by the homopolar mode were assumed to be governed by a short-range scattering constant,\cite{Fivaz_PR67,Song_PRL13} and DFT calculations show that this interaction is weaker than the Fr\"{o}hlich one.\cite{Sohier_PRB16} However, similar to the case of LO phonons, thickness fluctuations can generate an in-plane electric field, and this fact is often overlooked.

In the homopolar-mode case, the in-plane electric field in the mid-plane of the ML is governed by the negative charge of the chalcogen atomic sheets. That is, a charge is transferred in equilibrium conditions from the mid-plane (transition-metal atoms) to the top and bottom ones (chalcogen atoms).\cite{Ataca_JPCC12} This charge transfer is not changed by the atomic displacements. The curvature of the statically-charged chalcogen atomic sheets when the wavevector of the homopolar phonon is nonzero, as shown in Fig.~\ref{fig:zolo}(b),  introduces an in-plane electric field with which conduction-band electrons and valence-band holes can interact. The amplitude of the field is represented in Fig.~\ref{fig:zolo} by the color intensity of the `green clouds' in the mid-plane of the ML. This interaction is independent of the short-range interaction due to the volume change of the unit-cell. Note that the static charge on the chalcogen atomic sheets is not related to the effective Born charge that governs the Fr\"{o}hlich interaction with LO phonons. The latter is a dynamical effect that stems from the linear relation between the force on an atom and the macroscopic electric field.\cite{Gonze_PRB97}  

The in-plane electric fields causued by the homopolar and LO phonons can couple to each other. That is, homopolar phonons at certain wavelengths cannot be excited without provoking the LO phonons and vice versa; the two modes behave as a coupled system of charged harmonic oscillators (in the lowest order). Thus, the interactions between electrons and homopolar phonons or between electrons and LO phonons are not independent, and this property can explain the experimental result (the phonon mode with lower energy dominates the phonon-assisted recombination process). Furthermore, the mechanism can be extended to other layered materials. For example, the long-range ZO interaction should also exist in oxide-layered systems (e.g. cuprates), where the static charge transfer to the oxygen atoms is even larger than in ML-TMDs and where the transport takes place in atomic planes between those of the oxygen. 

\section{Identifying the signature of localization in the emission spectrum} \label{sec:discussion}

We discuss several localization-induced phenomena that manifest in the emission spectra of ML-TMDs. The sub-quadratic dependence of the five-particle peak on the excitation light intensity is addressed first.\cite{Chen_NatComm18,Ye_NatComm18,Li_NatComm18,Barbone_NatComm18} We then discuss experimental ways to tell apart recombination processes that involve localization effects or phonon-assisted transitions. In addition, we discuss the emission processes that result from the capture of dark and indirect excitons. Finally, we discuss the case of biexcitons and five-particle complexes.

\subsection{Amplitude dependence on the five-particle peak on excitation light intensity} \label{sec:subquad}
The densities of excitons have linear dependence on excitation light intensity regardless of whether they are direct, indirect, or dark. The same holds for trions because they are made of an electron (or hole) in the ML and an exciton. Consequently, it is expected that the amplitude of the biexciton peak should show quadratic dependence on excitation light intensity, and the same is expected if a delocalized five particle complex is made of an exciton and trion components. Nonetheless, experiments show that the amplitude of the five-particle peak has a  sub-quadratic  power law.\cite{Chen_NatComm18,Ye_NatComm18,Li_NatComm18,Barbone_NatComm18}  

The phonon-assisted recombination mechanism supports the observed behavior because the density of localized electrons that mediate this process decreases  when the excitation light  intensity increases. In addition to possible heating-induced decrease in their density (thermal activation), more and more localized electrons capture excitons and turn to localized trions when the density of excitons increases. When this capture happens and a trion instead of electron occupies the localization center, the biexciton cannot be captured. Namely, there are not enough quantum numbers to accommodate three holes (two from the biexciton and one from the negative trion) by states of the top two valleys in the valence-band. Furthermore, the two opposite-spin electrons of the localized negative trion are each valley-mixed and cannot accommodate any extra electron. All in all, while the density of biexcitons has quadratic dependence on excitation light intensity, the amplitude of the emitted light due to their phonon-assisted recombination becomes sub-quadratic when the density of localized electrons start to decrease.

\subsection{Telling apart recombination processes that involve localized and delocalized trions}

The observation of trions in the emission spectrum of ML-TMDs is ubiquitous. More often than not, the question whether localized or delocalized trions give rise to the emission is ignored. The answer to this question is not trivial because localized and delocalized trions appear in similar spectral positions, rendering it difficult to tell them apart. The reason for the spectral proximity is that the binding  energies of trions and electrons to the localization center are nearly the same (i.e., the exciton component of the trion contributes far less to the binding with the remote impurity). Therefore, the overall energy of a localized trion is nearly that of a delocalized trion plus the binding energy of a localized electron. Upon recombination we are still left with the localized electron, so that the energy of the emitted photon is similar to the one generated when a delocalized trion recombines and leaves behind a free electron.

Given the spectral proximity of the emitted photons from localized and delocalized trions, one can determine which type of trions recombine by comparing the PL spectrum to the one from differential reflectance spectroscopy. One can say that the emission is governed by localized trions if their optical transition is observed in the PL but not in differential reflectivity (under similar temperature and gate voltage).  The reason is that the latter measures the intrinsic absorption of the ML with equal contributions to the optical transition from the entire illuminated region. On the other hand, emission processes are influenced by  localized electrons or holes that capture thermalized excitons before they enter the minuscule light cone. 

The recombination of localized trions can be seen in PL experiments even when resident electrons or holes are depleted by surface residues or by a gate voltage that tunes the Fermi level to the energy gap. In this case, electrons and holes are re-introduced through the photoexcitation due to  dissociation of hot excitons. For example, if excitons are generated above the ground-state energy of the bright exciton or when exciton-exciton interaction annihilates one exciton and generates a hot electron and hole.\cite{Kumar_PRB14,Yuana_NS15,Yu_PRB16,Hoshi_PRB17,Lee_ACSP18,HanPRX_2018} Energy-relaxed electrons and holes can spend time in localization centers and capture thermal excitons. 

The fine-structure of negative trions in ML-WSe$_2$ or ML-WS$_2$ is an additional probe to identify if trions are delocalized. The relation between the valley and spin quantum numbers in the conduction band of these MLs  is such that the two electrons of a negative trion state can have singlet or triplet configurations.\cite{Jones_NatPhys16,Plechinger_NatCom16,Plechinger_NanoLett16,Courtade_PRB17} On the other hand, a localized trion can only have a singlet configuration because of the mixing between time-reversed states of the  conduction band (shutting out the valley quantum number). This difference between localized and delocalized negative trions in ML-WSe$_2$ (or ML-WS$_2$) means that one can observe a transition between localized-trion emission (one peak) to delocalized-trion emission (two peaks) when the gate-induced electrostatic doping changes from  neutral to electron-rich. The transition from one to two peaks in the spectrum takes place when the incipient density of delocalized electrons screens the attraction of excitons to localized electrons.

\subsection{Telling apart phonon-assisted recombination of excitons from recombination of localized trions}

Looking back at the short history of ML-TMDs, the signature of phonons in the emission spectrum was already present in the first report of trions in these materials.  Inspection of the gate-voltage-dependent PL measurements in the right panels of Fig.~2(a) in Ref.~[\onlinecite{Mak_NatMater13}] shows that the emission spectra can be fitted by two Lorentzians whose energy separation is nearly 50~meV. Indeed, the LO phonon mode in ML-MoS$_2$ is 48 meV and the homopolar  one is 51~meV.\cite{VanTuan_arXiv19,MolinaSanchez_PRB11} On the other hand, the gate-voltage-dependent absorption measurements, as shown in the same figure of Ref.~[\onlinecite{Mak_NatMater13}], reveal that the binding energy of delocalized trion with respect to a free electron and an exciton is evidently smaller. 

This example points to a simple test one can use. When the energy difference between the peaks of the exciton and three-particle complexes in the PL spectrum matches the phonon energy, then the emission process is likely phonon-assisted recombination of an exciton mediated by a localized electron. In this case, the capture is an integral part of the recombination process during which the localized trion is a virtual intermediate state.  When the energy difference does not match the phonon energy, then the recombination starts with a regular localized trion state and no phonons are involved in the radiative process. That is,  the recombination of the localized trion carries no information on the capture mechanism through which the trion was initially created. 

We conclude this part by mentioning two points regarding the nature of the phonon-assisted recombination process in ML-TMDs. The first one deals with the seemingly strong signature of this process in ML-WSe$_2$.\cite{Courtade_PRB17,Chen_NatComm18,Ye_NatComm18,Li_NatComm18,Barbone_NatComm18} A possible explanation can be that the energies of the LO and homopolar phonon modes are nearly degenerate in this material,\cite{Sahin_PRB13} rendering a strong coupling between them (Sec.~\ref{sec:zolo}). A recent experiment has confirmed the strong coupling of light with these two phonon modes in ML-WSe$_2$ through femtosecond surface x-ray diffraction.\cite{Tung_arXiv19} Similarly, long-lived homopolar phonon oscillations were observed in time-resolved transmission measurements.\cite{Jeong_ACSNano16} The second point deals with a possible enhancement of the phonon-assisted recombination process when neutral excitons become localized and not only charged particles or complexes. Exciton localization is likely to happen in defect-rich MLs and/or when charged impurities at the surface of the ML are abundant. When such localization is relevant, the wavefunction of the exciton is spread in momentum space and consequently its spontaneous radiative rate is suppressed.\cite{Wang_PRB16} At the same time, the capture matrix element is amplified due to a tighter overlap between the wavefunctions of the electron, exciton and trion states [Eq.~(\ref{eq:M3})]. Yet, creating real localized trion states is delayed when the phonon energy does not match the gained energy from the binding of an electron and exciton to form a localized trion. The reason is that  none of the involved parties has kinetic energy to offset the energy mismatch, and this fact restricts the ability to conserve energy. Under these conditions, the amplified matrix element mostly enhances the phonon-assisted recombination of excitons through generation of intermediate virtual trion states.

\subsection{Turning indirect or dark complexes to optically active}  

The emission spectra of ML-WS$_2$ and ML-WSe$_2$ include few peaks whose spectral position is 50-80 meV below that of the neutral exciton.\cite{Jadczak_nanotech17,Shang_ACSNano15,Ye_NatComm18,Smolenski_PRX16,Jadczak_NatComm19}  These optical transitions do not appear in the differential reflectance measurements suggesting the fingerprint of localization. Smole\'{n}ski \textit{et al.} have suggested that these peak emerge from trapping effects that involve dark excitons.\cite{Smolenski_PRX16} Dark excitons in tungsten-based MLs have considerable lower energy than bright ones ($\sim$40 meV in ML-WSe$_2$ and $\sim55$~meV in ML-WS$_2$).\cite{Wang_PRL17} The same is true for indirect excitons in these MLs,\cite{Dery_PRB15} indicating that  bright excitons relax to dark or indirect excitons. Considering the long recombination time of non-bright excitons,\cite{Tang_arXiv19} their densities should be evident in tungsten-based compounds compared with molybdenum-based ones in which bright excitons are the ground state.

\begin{figure}
 \centering
\includegraphics[width=8.5 cm]{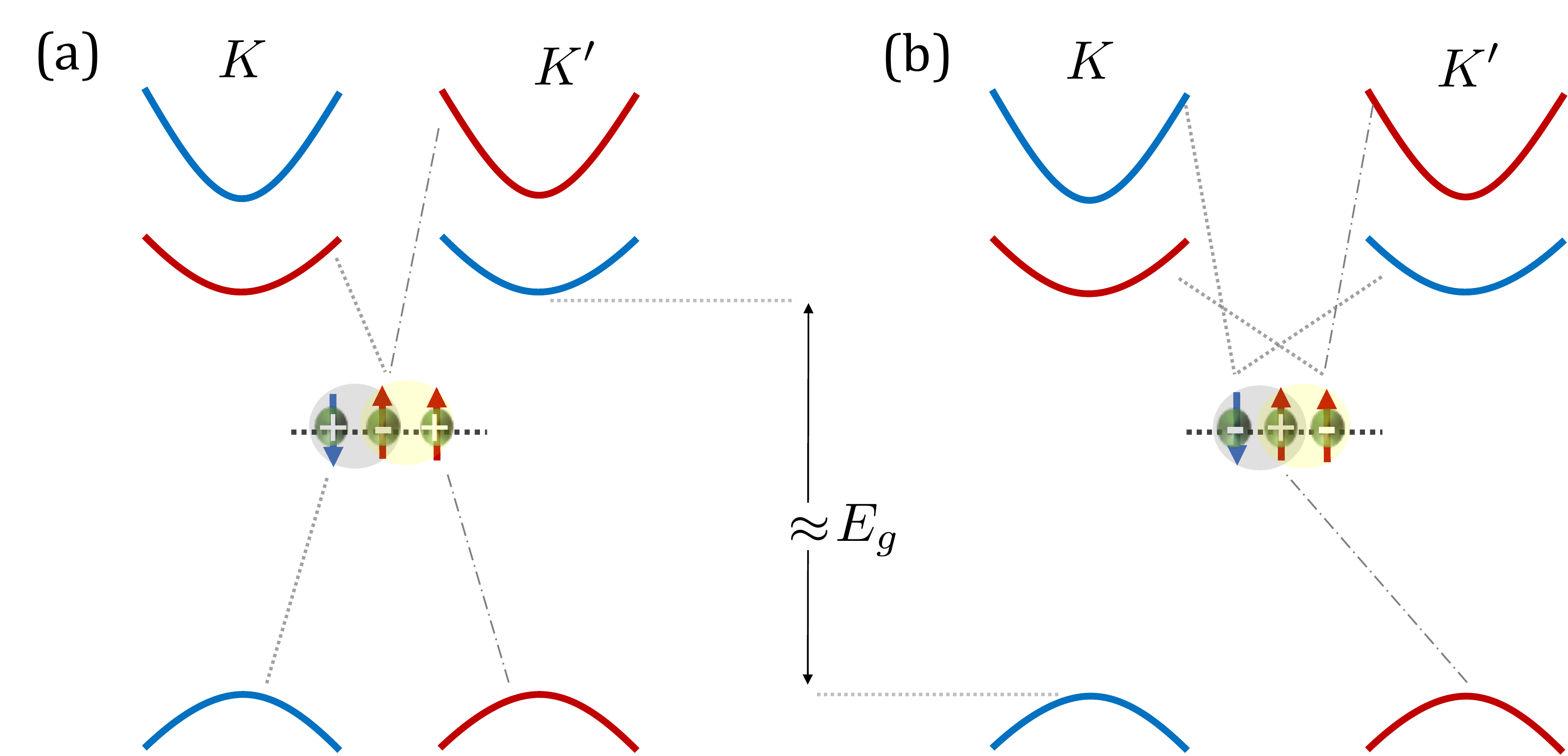}
 \caption{(a) A localized positive trion following the capture of a dark exciton by a localized hole. After the capture, the valley-mixing of the electron enables its radiative recombination with one of the two holes. The bright (dark) component is denoted by states from valleys connected to the particles by dashed-dotted (dotted) lines.  (b)  A localized negative trion following the capture of a dark exciton by a localized electron. After the capture, the valley-mixed electrons have components from all valley states.}\label{fig:loc_trion_from_dark}
\end{figure}

The confluence of  the localization-induced valley mixing and long lifetime of non-bright exciton population in tungsten-based MLs help us to better understand the observed behavior. When a dark or indirect exciton is captured by a localized electron (or hole), it becomes bright because of the valley mixing in the conduction band. Consider first a positive  trion where the localized hole captures the dark exciton. The electron component of the localized trion, as shown in Fig.~\ref{fig:loc_trion_from_dark}(a), becomes valley mixed and this effect brightens the trion. In the case of a negative  trion, the localized electron captures the dark exciton. Due to valley mixing, the two electrons of the localized trion involve components from all valleys of the conduction band, as shown in Fig.~\ref{fig:loc_trion_from_dark}(b). Emission from these positive and negative localized trions should be identified in the ballpark of 20 to 30 meV below the dark exciton peak in tungsten-based MLs (similar to the binding energies of trions with respect to bright excitons). This spectral region is the one in which the localization-related peaks are seen in the PL of tungsten- but not molybdenum-based MLs. A similar brightening effect can take place when a  negative dark trion is captured by a localized hole (or when a localized electron captures a positive dark trion). The only requirement is that same-charge particles have opposite spins. 

An additional interesting effect found by Smole\'{n}ski \textit{et al.} is that following excitation by a circularly polarized light, the circular polarization degree from the localization-related emitted photons  is strongly enhanced by applying a small out-of-plane magnetic field.\cite{Smolenski_PRX16} This behavior is consistent with the fact that the valley is no longer a good quantum number. In more detail, the spins of delocalized electrons or holes are robustly fixed in the out-of-plane direction due to the effective crystal magnetic field which has opposite directions in the $K$- and $K$'-point valleys. A localized electron, on the other hand, has contributions from both valleys and this intrinsic field is quenched. What is left are small internal effective magnetic fields due to exchange interaction between nearby localized electrons or hyperfine interactions that stem from defects in the ML, surface admolecules, or nearby oxygen vacancies and dangling bonds at the surface of SiO$_2$.\cite{Song_PRL14} The spin dephasing of localized electrons is induced by spin precession around the in-plane component of the tiny internal fields. Considering recombination of the positive trion complex [Fig.~\ref{fig:loc_trion_from_dark}(a)], the helicity of the emitted photon can be either $\sigma^+$ or $\sigma^-$ if the recombination rate of localized trions is slower than the Larmor frequency of the localized electron in the complex. One can suppress the spin precession and restore the circular polarization by applying an out-of-plane magnetic field that overcomes the small in-plane component of the internal fields.\cite{Song_PRL14}  

The PL experiments of ML-WS$_2$ and ML-WSe$_2$  show a series of peaks attributed to localization effects rather than one. Clearly, the microscopic origin of these peaks cannot all be attributed to localized positive trions due to negatively-charged remote impurities. For example, some of these defects may be attributed to tungsten vacancies in the ML that behave as acceptors,\cite{Zhang_PRL17,Edelberg_arXiv18} or to oxygen atoms that substitute Se atoms or positioned in interstitial sites (both take electrons from the crystal).\cite{Zheng_arXiv19} The short-range potential of such defect centers can localize holes, which in turn trap excitons and form localized positive trions. In addition, it remains to be seen whether the spin of a localized hole (or of the hole in a localized negative trion) can lose its valley identity and precess (see Sec.~\ref{sec:holes}). When this case is applicable, then the localization-related peaks seen in the PL spectrum can be attributed not only to localized positive trions but to localized excitons and negative trions as well. Otherwise, transmission electron microscopy and similar spectroscopies should be able to reveal the other types of acceptor defects. Along with ab-initio modeling of the revealed defects and their hyperfine fields, one can reinforce or refute our proposed  physical picture.

\subsection{Localized vs delocalized charged biexcitons}

Delocalized biexciton and five-particle complexes have been observed in ML-MoSe$_2$ using non-linear coherent spectroscopy experiments.~\cite{Hao_NatCommun17} The ultrafast nature of such wave-mixing spectroscopy implies that the neutral and charged biexciton features are genuine delocalized complexes. The use of mixed waves with opposite circular polarization creates biexciton states from two bright-exciton components (one from each valley). Contrary to this case,  the biexciton is made of dark and bright exciton components in PL experiments where both neutral and charged biexcitons are observed.\cite{Chen_NatComm18,Ye_NatComm18,Li_NatComm18,Barbone_NatComm18} Namely, the two electrons of the biexciton have the same spin but opposite valleys. Biexcitons of such nature cannot be observed in differential reflectance spectroscopy because bright excitons turn dark or indirect during their thermalization and not at the instance of photexcitation.

The following two arguments are used to reinforce the claim that the observed five-particle complexes in PL experiments involves localization. The first argument is that a localized electron can only interact with biexcitons made of bright and dark exciton components, as shown in Fig.~\ref{fig:diagram}(c). This is the type of biexcitons seen in PL experiments. Conversely, if biexcitons made of two bright-exciton components are to be detected in PL experiments, then the charged state of such biexciton is delocalized in nature (there are not enough quantum numbers for this biexciton to interact with a valley-mixed localized electron).  The second argument relies on the comparison between the binding energies of the neutral and charged biexcitons measured in PL and non-linear coherent spectroscopy experiments. The latter shows that the binding energy of the biexciton in ML-MoSe$_2$ is around 20~meV whereas that of the five-particle complex is $\sim$5~meV with respect to the trion (and only about 15-20~meV below the biexciton peak).\cite{Hao_NatCommun17} While PL experiments of ML-WSe$_2$ find a similar binding energy for the biexciton, they show a much different binding for the five-particle complex.\cite{Chen_NatComm18,Ye_NatComm18,Li_NatComm18,Barbone_NatComm18} 

\section{CONCLUSIONS and oulook} \label{sec:conclusions}

This work has highlighted several important effects of charge localization on the optical properties of ML-TMDs.  We have analyzed the valley mixing of localized electrons and the type of exciton complexes that can interact with a localized electron. We have then reviewed the findings of recent experiments that probed biexcitons and five-particle complexes in the PL of ML-WSe$_2$. The origin for the five-particle complex was studied by analyzing two possible scenarios: recombination of a delocalized complex or phonon-assisted recombination of a biexciton mediated by a localized electron. A stronger support for the latter mechanism was found, and we have analyzed its microscopic details as well as dependence on magnetic field.  An intriguing coupling between longitudinal-optical and homopolar phonon modes in ML-TMDs was discussed. This coupling can be the reason that the phonon with lower energy of the two modes dominates the Raman spectra and phonon-assisted recombination. Finally,  ways to identify recombination processes that involve localized complexes or  optical phonons were discussed. We have discussed the brightening of dark and indirect excitons following their capture by localized holes or electrons (this scenario is applicable in tungsten-based MLs where the densities of dark and indirect excitons is non-negligible). 

Hopefully, this work would motivate further studies of the relation between localization and emission processes in ML-TMDs. For example, exciton-exciton annihilation is a detrimental non-radiative recombination process that limits the quantum yield in these materials.\cite{Kumar_PRB14,HanPRX_2018} Experiments show that it becomes the dominant recombination mechanism already when the  exciton density is $\sim$10$^{10}$~cm$^2$.\cite{Kumar_PRB14,Yuana_NS15,Yu_PRB16,Hoshi_PRB17,Lee_ACSP18} If we ignore localization effects, then the average distance between excitons is $\sim$100~nm at these densities, which is two orders of magnitude larger than the exciton radius in ML-TMDs.\cite{Stier_NanoLett16} Given that the dipole-dipole interaction that glues excitons decays as fast as $\sim 1/r^6$ with the distance $r$ between excitons, it is hard to imagine how the exciton-exciton annihilation or the formation of biexcitons is effective unless the exciton density is markedly large.

The scenario changes in the presence of localized electrons because they attract excitons, and therefore, locally increase the exciton density. As such, localized electrons enhance the exciton-exciton scattering rates as well as the formation of biexcitons. One can test this hypothesis by employing various ways to fabricate devices. For example, one can control the defect concentration at the surface, or deposit the stack of hBN/ML-TMD/hBN on top of passivated SiO$_2$,\cite{Ajayi_2DMater17,Tinkey_APL16} or possibly use substrates with higher dielectric constants to suppress the role of remote defects. 

In closing, compared with the impressive understanding of the relation between defects and optical properties in typical semiconductors,\cite{Walle_JAP04,Pantelides_RMP78,Cardona_RMP05} it is clear that extensive efforts are needed before we can reasonably characterize defect-induced optical properties of ML-TMDs. In view of the large surface to volume ratio of atomically-thin semiconductors, understanding the effect of remote  and  internal impurities is imperative for the understanding of emission processes. 

\acknowledgments{This work was supported by the Department of Energy, Basic Energy Sciences (Grant No. DE-SC0014349). The computation work (SVM of few particle complexes) was also supported by National Science Foundation under Contract No. DMR-1503601.}

\appendix

\begin{figure*}
\includegraphics[width=15.0cm]{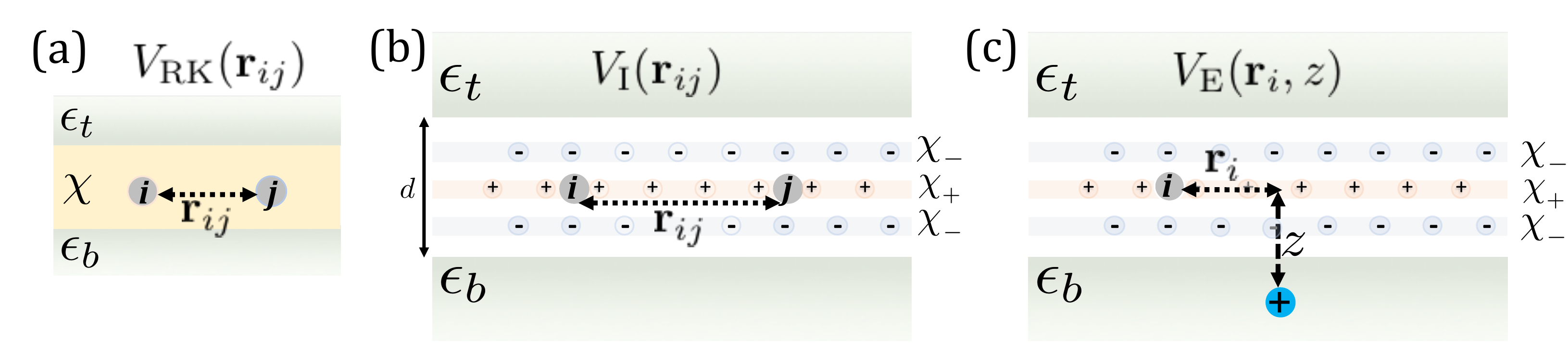}
 \caption{(a)  The dielectric environment when considering a uniform ML with polarizability $\chi$. The solution of the Poisson Equation in the ideal 2D limit (zero thickness of the ML) leads to the celebrated Rytova-Keldysh potential, $V_{\text{RK}}(\mathbf{r}_{ij})$ between the charges $i$ and $j$. (b) The revised ML geometry when modeling the Coulomb interaction between the charged particles in the mid-plane of a ML of thickness $d$. The ML is viewed as three atomic sheets with polarizabilities $\chi_+$ for the central one (Mo/W) and $\chi_-$ for the top and bottom ones (S/Se/Te, displaced by $\pm$d/4 from the center).  Screening from the chalcogen  sheets helps to confine the field lines in the ML, thereby reducing the dependence on the bottom and top materials whose dielectric constants are $\epsilon_b\,\&\,\epsilon_t$. (c) The same as in (b) but when modeling the Coulomb interaction between a charged particle in the mid-plane of a ML and a remote positively-charged defect embedded in the bottom dielectric layer. }
  \label{fig:3chi}
\end{figure*}

\section{Coulomb potentials} \label{app:V}
The Coulomb interactions between charged particles in the ML, and between the remote charged impurity and charged particles are analyzed in this section. We start with the former.

\begin{figure*}
\includegraphics[width=12.0cm]{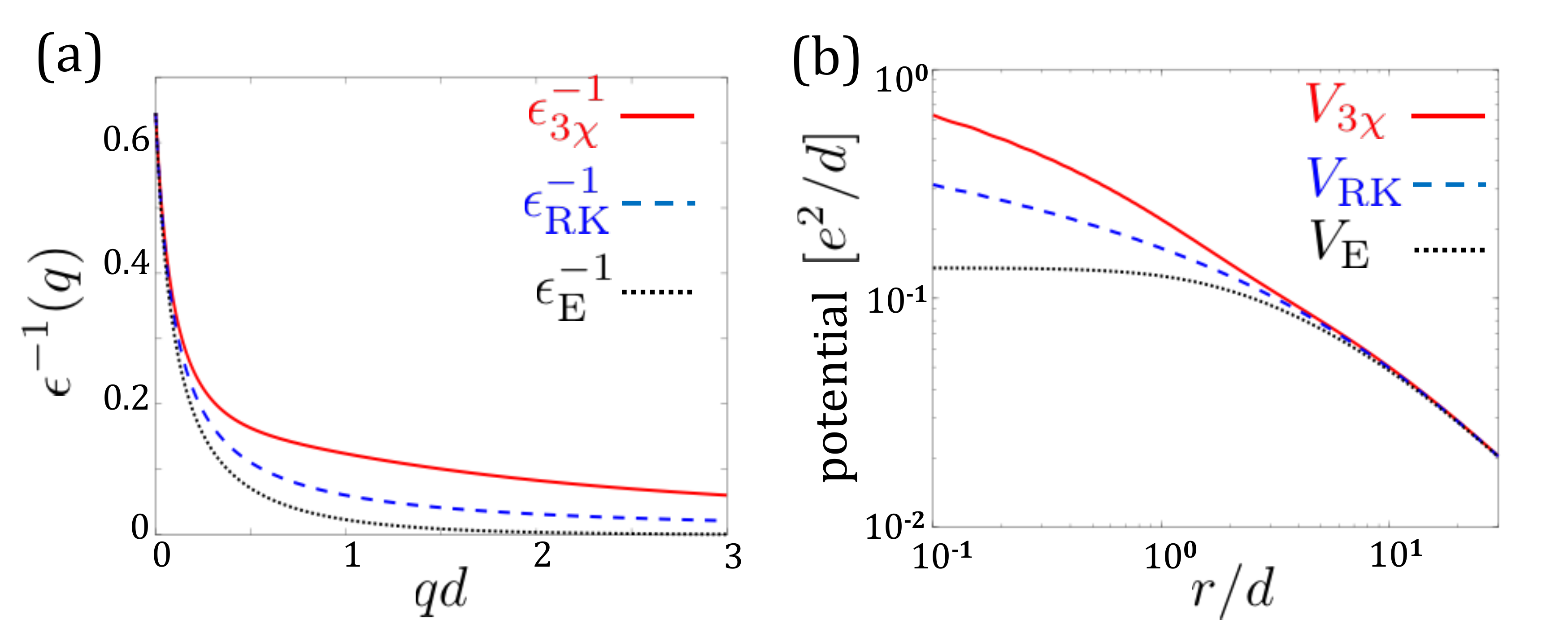}
 \caption{(a)  Inverse of the static dielectric functions, following Eqs. (\ref{Eps_RK}), (\ref{EpsI}), and (\ref{Eq:BareCoulomb}). (b) The resulting potential forms in real space. The parameters we use are $d=0.6$~nm, $\epsilon_t=1$ (air), $\epsilon_b=2.1$ (SiO$_2$), $\ell_+=\ell_-=5d$, $r_0=15d$. For the remote impurity case (dotted black lines), we assume a defect at the surface of SiO$_2$, $z=-0.5d$. } \label{fig:eps_pot}
\end{figure*}

Regardless of the model we use for our 2D system, the solution of the Poisson Equation leads to the following 2D Fourier transform of the Coulomb potential,\cite{VanTuan_PRB18,Cudazzo_PRB11},
\begin{equation}
 V_I({ q})=\frac{2\pi e_ie_j}{A\epsilon_\text{I}(q)q}\,\,, \label{eq:2D_potential_Fourier}
 \end{equation}
where $e_i$ and $e_j$ are the charges of the interacting particles, and $A$ is the area of the ML. The static dielectric function, $\epsilon_\text{I}({q})$, depends on the model we use. For the geometry in Fig.~\ref{fig:3chi}(a), we get the Rytova-Keldysh potential\cite{Rytova_MSU67,Keldysh_JETP79,Cudazzo_PRB11}
\begin{equation}
 \epsilon_\text{I}({\bf q}) =   \epsilon_\text{RK}({\bf q}) = \frac{\epsilon_b + \epsilon_t}{2} + r_0 q, \label{Eps_RK}
 \end{equation}
where $\epsilon_{b(t)}$ is the dielectric constant of the bottom (top) dielectric layer, and $r_0=2\pi\chi$ is the screening length in the ML due to its polarizability, $\chi$. If we use the  geometry shown in Fig.~\ref{fig:3chi}(b), we get\cite{VanTuan_PRB18}
\begin{equation}\label{EpsI}
\epsilon_{\text{I}}(q) =   \epsilon_{3\chi}({\bf q}) =\frac{1}{2}\left[\frac{N_t(q)}{D_t(q)}+\frac{N_b(q)}{D_b(q)}\right],  
\end{equation}
where
\begin{widetext}
\begin{eqnarray}
D_\gamma(q) &=& 1+q\ell_- -q\ell_- (1+p_\gamma)\text{e}^{-\frac{qd}{2}} - (1-q \ell_- ) p_\gamma \text{e}^{-qd}, \nonumber \\
N_\gamma(q) &=& \left(1+q\ell_-\right)\left(1+q\ell_+\right) \nonumber  +  \left[\left(1-p_\gamma\right)-\left(1+p_\gamma\right)q\ell_+\right]q\ell_-\text{e}^{-\frac{qd}{2}}  + (1-q\ell_-)(1-q\ell_+ )p_\gamma\text{e}^{-qd}. \label{eq:DN}
\end{eqnarray}
\end{widetext}
$d$ is the thickness of the ML, $p_\gamma \equiv (\epsilon_\gamma-1)/(\epsilon_\gamma+1)$ for $\gamma=\{b,t\}$, and $ \ell_\pm=2\pi\chi_\pm$ are the dielectric screening lengths of each of the atomic sheets in the ML, as shown in  Fig.~\ref{fig:3chi}(b). Note that $\epsilon_{3\chi}({\bf q}) = \epsilon_{\text{RK}}({\bf q}) $ when $d=0$ and $r_0 = \ell_+ + 2\ell_-$. Figure~\ref{fig:eps_pot}(a) shows the inverse dielectric functions  for a ML supported on SiO$_2$ and exposed to air, $\epsilon_b=2.1$ and $\epsilon_t=1$. The solid (dashed) line shows the 3$\chi$ (Rytova-Keldysh) model. Their values converge to $2/(\epsilon_b+\epsilon_t)$ when $q \rightarrow 0$. 

The real-space Coulomb potential then follows from
\begin{equation}
V_{\text{I}}({r}) =  e_ie_j \int_0^{\infty}  \! \! dq \frac{ J_0(qr)}{\epsilon_{\text{I}}(q)} \,\, , \,\,\,\, \label{eq:2D_real}
 \end{equation}
where $J_0$ is the zeroth-order Bessel function. Figure~\ref{fig:eps_pot}(b) shows the real-space potential forms using the dielectric static functions in Figure~\ref{fig:eps_pot}(a). $V_{\text{RK}}$ and $V_{3\chi}$ converge to the conventional potential $2e_ie_j/(\epsilon_b+\epsilon_t)r$ when $r > \{d,r_0,\ell_{\pm}\}$. Their values deviate at short distances, $r \lesssim d$, corresponding to the region $1  \lesssim qd$ in  Fig.~\ref{fig:eps_pot}(a).

\subsection*{Coulomb interaction between a particle in the mid-pane of the ML and a remote defect, $V_{\text{E}}$}

Repeating the steps used to derive $V_{3\chi}$,\cite{VanTuan_PRB18} we can similarly derive the Coulomb potential between a charged particle in the mid-plane of the ML ($z'=0$) and a remote charged defect embedded in the bottom (or top) dielectric layers.  Its solution comes from the relation, $V(q,z)= e_i \phi_{\bm{q}}(z'=0,z)$, where $\phi_{\bm{q}}(z'=0,z)$ is the 2D Fourier transform of the remote-defect-induced potential. The Poisson Equation in this case reads
\begin{eqnarray}
&& \frac{\partial}{\partial z'}\!\left[\!\kappa(z')\frac{\partial\phi_{\bm{q}}(z',z)}{\partial z'}\!\right]\!-\kappa(z')q^2\phi_{\bm{q}}(z',z)  -\frac{4\pi e_d}{A}\delta\left(\!z'\!-\!z \right) \nonumber \\
&+&  2 q^2 \! \left[\delta(z')\ell_+ + \delta\!\left(\!z'\!-\!\tfrac{d}{4}\right)\!\ell_- + \delta\!\left(\!z'\!+\!\tfrac{d}{4}\right)\!\ell_- \! \right]\!\phi_{\bm{q}}(z',z), \label{Eq:Poisson_Ext}
\end{eqnarray}
where $e_d$ denotes the charge of the defect, $z$ its distance from the mid-plane of the ML, and
\begin{equation}\label{Eq:DielEnv2}
\kappa(z')=\left\{\begin{array}{ll}
 \epsilon_t & \mathrm{for}\quad z'>d/2,\\
 1 & \mathrm{for}\quad -d/2<z'<d/2,\\
 \epsilon_b & \mathrm{for}\quad z'<-d/2\,.
 \end{array}\right.
\end{equation}
Fixing the point-charge defect to the bottom layer, $z < -d/2$,  one can solve Eq.~(\ref{Eq:Poisson_Ext}) with the boundary conditions that $\phi_{\bm{q}}(z',z)$ is continuous and its derivative is piecewise continuous with jumps of $2q^2\ell_+\phi_{\bm{q}}(0,z)$ at $z'=0$, $2q^2\ell_- \phi_{\bm{q}}(\pm d/4,x)$ at $z'=\pm d/4$, and $-4\pi e_d/A$ at $z'=z$. The Coulomb interaction between $e_d$ and a charged-particle $e_i$ in the mid-plane of the ML yields  
\begin{eqnarray}
\!\!\!  \!\!\!  V(q,z) &=& e_i \phi_{\bm{q}}(z'=0,z) = \frac{2\pi e_de_i}{A\epsilon_{\text{E}}(q,z)q}\,\,, \\  \!\!\!  \!\!\!\!\!\! \epsilon_{\text{E}}^{-1}({\bf q},z)&=&\frac{2\left(c_0 + c_1 e^{\frac{qd}{2}} + c_2 e^{qd}\right)  e^{q(z+d)} }{ d_0 + d_1 e^{\frac{qd}{2}} + d_2 e^{qd}+ d_3 e^{\frac{3qd}{2}} + d_4 e^{2qd}} \,,  \label{Eq:BareCoulomb} \,\,\,
\end{eqnarray}
where 
\begin{eqnarray}
c_0&=& \left(q\ell_- -1\right) \left(\epsilon_t-1\right),  \nonumber  \\
c_1&=& -2 q\ell_- \epsilon_t , \nonumber \\
c_2&=&  \left(q\ell_- +1\right) \left(\epsilon_t+1\right),
\end{eqnarray}
and
\begin{widetext}
\begin{eqnarray}
d_0&=&  \left(q\ell_- -1\right){}^2 \left(q\ell_+-1\right) \left(\epsilon_t-1\right) \left(\epsilon_b-1\right) ,\nonumber \\
d_1&=& -2q\ell_- \left(q\ell_- -1\right)  \left[1-q\ell_+(\epsilon_t+ \epsilon_b) +  \left(2q\ell_+ -1\right) \epsilon_t\epsilon _b \right],  \nonumber\\
d_2&=&  -2 q^2\ell_-^2 \left(\epsilon_t+\epsilon_b\right)+2 q\ell_+ \left[1-q^2\ell_-^2+\left(3 q^2\ell_-^2 -1 \right) \epsilon_t \epsilon_b\right], \nonumber  \\
d_3 &=& -2 q\ell_- \left(q\ell_- +1\right) \left[-1+q\ell_+(\epsilon_t+ \epsilon_b) + \left(2q\ell_+ +1\right) \epsilon_t\epsilon_b  \right], \nonumber \\
d_4&=& \left(q\ell_- +1\right){}^2 \left(q\ell_+ +1\right) \left(\epsilon_t+1\right) \left(\epsilon_b+1\right).
\end{eqnarray}
\end{widetext}
The dotted-black line in Fig.~\ref{fig:eps_pot}(a) shows the resulting inverse dielectric function for a surface defect, $z=-d/2$, with charge $e_d=e$. All other values are the same as before: $d=0.6~nm$, $\ell_{\pm}=5d$, $\epsilon_t=1$ and $\epsilon_b=2.1$. The ensuing real-space 2D interaction,  
\begin{eqnarray}
V_{\text{E}}(r,z) =  e_de_i \int_0^{\infty}  \! \! dq \frac{ J_0(qr)}{\epsilon_{\text{E}}(q,z)} \,\, , \,\,\,\, \label{eq:2D_potential_Fourier}
\end{eqnarray}
is shown by the dotted-black line in Fig.~\ref{fig:eps_pot}(b). The potential converges to the conventional one $2e_de_i/(\epsilon_b+\epsilon_t)r$ when $r > \{d,z,\ell_{\pm}\}$. Its value saturates at short distances, $r < z$, because the potential scales as $1/\sqrt{z^2+r^2}$.  

\section{Stochastic Variation Method}\label{app:SVM}
The wavefunctions of the delocalized biexciton, localized electron and localized charged-biexciton are expressed  by 
 \begin{eqnarray} 
\Psi_{XX^0}( \mathbf{K}) &=&     \frac{\exp({i\mathbf{K}{\mathbf{R}}})}{\sqrt{A}}\varphi_{XX^0}({\bf r}_1,{\bf r}_2,{\bf r}_3,{\bf r}_4), \nonumber \\ 
\Psi_{\ell}      &=&      \varphi_\ell({\bf r}_5), \nonumber \\ 
\Psi_{XX^-}      &=&        \varphi_{XX^-}({\bf r}_1,{\bf r}_2,{\bf r}_3,{\bf r}_4,{\bf r}_5) . \label{eq:wf_145}
\end{eqnarray} 
${\bf r}_1={\bf r}_{e_1},{\bf r}_2={\bf r}_{h_1},{\bf r}_3={\bf r}_{e_2},{\bf r}_4={\bf r}_{h_2}$ are the real-space coordinates of the biexciton's electrons and holes, while ${\bf r}_5={\bf r}_{\ell}$ is the coordinate of the localized electron. In addition, $A$ is the area of the ML, ${\bf R}=\sum_{i=1}^4 m_i {\bf r}_i/\sum_{i=1}^4 m_i$ is the Center of Mass (CoM) coordinate of the delocalized biexciton, and $\bf K$ is its wavevector. Using the SVM,\cite{Varga_CPC08,Mitroy_RMP13,Suzuki_Varga_Book98} the wavefunctions are  expressed as sums of correlated Gaussians, 
\begin{eqnarray} 
&& \varphi(\mathbf{r}_1, \mathbf{r}_2, ...., \mathbf{r}_{\mathcal{N}}) =   \sum_i^{n_{\mathcal{N}}} C_i \exp\left( -\frac{1}{2} {\bf r}^\text{T} \mathcal{A} \,\,\, {\bf r}  \right) \nonumber \\
  &\,\,\,\,\,& \,\,\,\,\,\,\,\,\,\,\,\, \equiv \sum_i^{n_{\mathcal{N}}} C_i \exp\left( -\frac{1}{2} \sum_{k<l}^{\mathcal{N}} \alpha_{kl}^i r_{kl}^2 -\frac{1}{2} \sum_{k}^{\mathcal{N}} \beta_{k}^i r_k^2   \right).         \,\,\,\,\,\,\,\,\,\,\,\,\label{eq:svm_form}
\end{eqnarray} 
$n_{\mathcal{N}}$ is the number of correlated Gaussians needed to describe the wavefunction of the $\mathcal{N}$-particle complex. Other parameters are $r_{kl} = |\mathbf{r}_k - \mathbf{r}_l |$ and ${\bf r}=\{ {\bf r}_1,{\bf r}_2,...,{\bf r}_N\}^\text{T}$, and the variational parameters are $C_i$, $\alpha_{kl}^i$ and $\beta_{k}^i$ where the latter two define the $\mathcal{N} \times N$ matrix $\mathcal{A}$. The variational parameters are found by minimizing the energy associated with the (Mott-Wannier) Hamiltonian of the $\mathcal{N}$-particle  complex
\begin{eqnarray}
H_{\mathcal{N}} =  \sum_i^{\mathcal{N}} V_{\text{E}}(r_{i},z) - \frac{\hbar^2}{2m_i}\nabla^2_i  + \sum_{i<j}^{\mathcal{N}} V_{\text{I}}(r_{ij})  \,.  \label{eq:H}
\end{eqnarray}
$V_{\text{E}}(r_{i},z)$ is the Coulomb interaction between the $i^{\text{th}}$ particle in the ML  and a remote charged defect whose distance from the mid-plane of the ML is $z$, as shown by Fig.~\ref{fig:localE}(a). $m_i$ is the effective mass of the $i^{\text{th}}$ particle and $V_{\text{I}}(r_{ij})$ is the Coulomb interaction between this particle and the $j^{\text{th}}$ one where $r_{ij}=|{\bf r}_i -{\bf r}_j|$. The latter interaction is relevant when $\mathcal{N} \geq 2$. Details on the Coulomb interactions were discussed in Appendix~\ref{app:V} and their parameters are summarized in Appendix~\ref{app:parameters}. 

\subsection{The Fr\"{o}hlich interaction}\label{app:frohlich}

The Fr\"{o}hlich coupling parameter in Eq.~(\ref{eq:M5}) reads\cite{Sohier_PRB16}
\begin{eqnarray}
D_{j,E_2'}(q) &=& \beta_j  \sqrt{n_{E_2'} + \frac{1}{2} \pm \frac{1}{2} } \sqrt{\frac{A_u}{A}}\sqrt{\frac{\hbar^2}{2M_xE_{E_2'}}}  \nonumber \\ && \,\,\left( 1 + \sqrt{\frac{M_x}{M_m} }\right) \frac{2\pi Z_{E_2'} e^2}{A_u\epsilon_{\text{I}}(q)}   \,\,\,,  \label{eq:Frohlich_coupling}
\end{eqnarray}
where $\beta_j=1\,(-1)$ when the $j^{\text{th}}$ particle is a hole (electron). $n_{E_2'} = 1/[\text{exp}(E_{E_2'}/k_BT)-1]$ is the Bose Einstein distribution. $E_{E_2'}$ is the phonon energy where we have neglected its weak dependence on $\mathbf{q}$ due to the dispersionless nature of long-wavelength optical phonons. The $\pm$ sign denotes phonon emission (plus) or absorption (minus), where only the former is relevant at low temperatures in which $n_{E_2'} \rightarrow 0$. $A$ and $A_u$ are the areas of the ML and unit cell, respectively.  $M_x$ and $M_m$ are the masses of the chalcogen and transition-metal atoms, respectively. $Z_{E_2'}$ is the Born effective charge describing the linear relation between the force on the atom and the macroscopic electric field. Conservation of charge implies that $Z_{E_2'}=Z_{m}=2Z_x$. $\epsilon_{\text{I}}(q)$ is the static dielectric function (Appendix~\ref{app:V}).

\subsection{The calculation of Eq.~(\ref{eq:M5})}\label{app:MK}

The calculation of the matrix-element in Eq.~(\ref{eq:M3}) was studied in Ref.~[\onlinecite{VanTuan_arXiv19}]. Here, we study the matrix element in Eq.~(\ref{eq:M5}) whose evaluation involves the following multivariable integration,
\begin{widetext}
\begin{eqnarray}
 F(\mathbf{K},\mathbf{q}) = \sum_{j=1}^5 \beta_j F_j(\mathbf{K},\mathbf{q}) =   \sum_{j=1}^5   \beta_j  \left\langle  \Psi_{XX^-}(\{ {\bf r} \})  \left| \exp \left( i\mathbf{q}\mathbf{r}_{j} \right)    \right| \Psi_{XX^0}({\bf r}_1,{\bf r}_2,{\bf r}_3,{\bf r}_4; \mathbf{K}) \Psi_{\ell}({\bf r}_5)   \right\rangle  .   \label{eq:FKq}
\end{eqnarray} 
\end{widetext}
$\beta_j=1\,(-1)$ if the $j^{\text{th}}$ particle in the complex is a hole (electron). The advantage of using the correlated Gaussians in Eq.~(\ref{eq:svm_form}) to evaluate Eq.~(\ref{eq:FKq}) is that the tedious integration over five two-dimensional coordinates can be carried analytically. Specifically, we have found that the localized electron state can be accurately expressed as the sum of a few tens correlated Gaussians ($n_1$), while the biexciton and localized charged-biexciton states by a few hundreds ($n_4$ and $n_5$). As a result, by substituting Eqs.~(\ref{eq:wf_145})-(\ref{eq:svm_form}) into  Eq.~(\ref{eq:M5}), the matrix element can be expressed as a sum over $n_1n_4n_5$ terms. Compared with the time it takes to find the variational parameters of these correlated Gaussians, the calculation of the matrix element is instantaneous. 

The integration in Eq.~(\ref{eq:FKq}) can be performed more easily upon the transformation, ${\bf x}=U {\bf r}$, of the  Cartesian coordinates ${\bf r}=\{ {\bf r}_1,{\bf r}_2,{\bf r}_3,{\bf r}_4,{\bf r}_5\}^\text{T}$ to Jacobi coordinates $\ {\bf x}=\{ {\bf x}_1,{\bf x}_2,{\bf x}_3,{\bf x}_4,{\bf x}_5\}^\text{T}$. The transform matrix is given by \cite{Mitroy_RMP13,Suzuki_Varga_Book98} 
\begin{equation}
U = 
 \begin{pmatrix}
  1 & -1 & 0 & 0  & 0 \\
  \frac{m_1}{m_{12}} & \frac{m_2}{m_{12}} & -1 & 0   &  0 \\
  \frac{m_1}{m_{123}} & \frac{m_2}{m_{123}} & \frac{m_3}{m_{123}} & -1   &  0 \\
  \frac{m_1}{m_{1234}} & \frac{m_2}{m_{1234}} & \frac{m_3}{m_{1234}} &  \frac{m_4}{m_{1234}}   &  -1 \\
  \frac{m_1}{m_{12345}} & \frac{m_2}{m_{12345}} & \frac{m_3}{m_{12345}} &  \frac{m_4}{m_{12345}}   &  \frac{m_5}{m_{12345}} 
 \end{pmatrix},
\end{equation}
where $m_{1\cdots M}=\sum_{j=1}^M m_j$. The CoM coordinate of the five-particle complex is the final component in the Jacobi coordinates system. The wavefunction of five-particle system becomes
\begin{equation}
 \Psi_{XX^-}\left( {\bf x}  \right) =\sum_{i=1}^{n_5} C_{5,i} e^{-\frac{1}{2}  {\bf x}^{\text{T}}A_{5,i}{\bf x}} ,
 \end{equation}
where $n_5$ is the number of correlated Gaussians needed to correctly describe the 5-particle complex. The matrices in Jacobi coordinates  can be obtained from those in Cartesian coordinates by 
\begin{equation}
   A_{5,i}=\left( U^{-1}\right)^{\text{T}} \mathcal{A}_{5,i} {U^{-1}}.
  \end{equation}
 Similarly, the wavefunction for four-particle (biexciton) and one-particle (localized electron) systems can also be rewritten as
\begin{equation}
 \Psi_{XX^0}({\bf x})=\frac{e^{i{\bf K}\,\, G \,\,U^{-1} {\bf x} }}{\sqrt{A}} \sum_j^{n_4} C_j^B e^{-\frac{1}{2}  {\bf x}^{\text{T}}A_{4,j}{\bf x}}  ,
 \end{equation}
and
\begin{equation}
 \Psi_{\ell}({\bf x})=  \sum_i^{n_1} C_i^\ell e^{-\frac{1}{2}  {\bf x}^{\text{T}}A_{1,i}{\bf x}}  ,
 \end{equation}
 where $G= \left(m_1,m_2,m_3,m_4,0  \right)/m_{1234}$ and ${\bf R}=G \,\,U^{-1} {\bf x}$ is the CoM coordinate of the biexciton. The overlap factor $F_\gamma(\mathbf{K},\mathbf{q})$ becomes 
\begin{eqnarray}
&& \!\!\!\!\!\!\!\!\!\!\!\! \!\!\!\!\!\! F_\gamma(\mathbf{K},\mathbf{q}) = \frac{1}{\sqrt{A}} \sum_{i,j,k}C_{5,i}C_{4,j}C_{1,k}  \int d{\bf x}     e^{-\frac{1}{2} {\bf x}^{\text{T}}A_{ijk}{\bf x}+v_\gamma^{\text{T}}{\bf x}}             \nonumber \\ 
  \,\,\,\,\,\,\,&=&  \frac{1}{\sqrt{A}} \sum_{i,j,k}C_{5,i}C_{4,j}C_{1,k}   \frac{\left( 2\pi\right)^5}{\text{det}\,\,(A_{ijk})}      e^{\frac{1}{2}v_\gamma^{\text{T}}A_{ijk}^{-1}v_\gamma} ,
\end{eqnarray}
 where $A_{ijk}=A_{5,i} + A_{4,j} +A_{1,k}$ and $v_\gamma^{\text{T}}=i\left( {\bf K} G +{\bf q}I_\gamma \right) \,\,U^{-1} $.  $I_\gamma$ is diagonal $5\times 5$ matrix with every element 0 except $I_{\gamma\gamma}=1$. 
 
\section{Parameters}\label{app:parameters}

The simulations are carried by considering the geometry in Fig.~\ref{fig:3chi}(b). The values we have used for the screening parameters in the Coulomb potentials are $\ell_+ = \ell_- =5.9d$ (with $d=0.6$~nm). When the ML is supported on SiO$_2$ and exposed to air, we have used $\epsilon_b=2.1$ and $\epsilon_t=1$. We have used the high-frequency limit for the dielectric constant of SiO$_2$ because of the non-negligible binding energy of the localized electron (i.e., the atoms in SiO$_2$ cannot trace the fast motion of the localized electron in the ML around the impurity center). When the ML is encapsulated in hBN, we have used $\epsilon_b=\epsilon_t=3.8$.\cite{VanTuan_PRB18}

The effective mass values are considered as follows. For neutral complexes in ML-WSe$_2$, we have used the bare effective masses at the band edges: $m_{cb} = 0.4m_0$ in the conduction-band bottom valley, $m_{ct} = 0.29m_0$ in the conduction-band top valley, and $m_{vt} = 0.36m_0$ for the hole in the valence-band top valley.\cite{Kormanyos_2DMater15}. When dealing with the localized electron,  we have added the polaron effect according to $m_\ell = (m_{ct} + m_{cb})*(1+ \alpha_P)/2$. The bare mass is given in terms of the average between the effective masses of the top and bottom valley (valley mixing) and the polaron parameter is $\alpha_P=0.34$.\cite{VanTuan_PRB18}  In case of delocalized charged complexes (Table~\ref{tab:E_deloc}), the polaron effect was equally distributed among the two electrons of the delocalized dark trion, $m_{cb}*(1+ 0.5\alpha_P)$. Similarly, in case of the delocalized five-particle complex (made of dark trion and bright exciton components), the distributed polaron effect follows $m_{cb}*(1+ 0.33\alpha_P)$ for two of its electrons and $m_{ct}*(1+ 0.33\alpha_P)$ for the third electron.

Using the above dielectric screening parameters and effective masses, we have achieved very good agreement between the calculated binding energies and empirical values of delocalized complexes  (see Appendix~\ref{app:benchmark_delocalized}). When calculating the capture process, we have assumed that the phonon energy is $E_{E_2'}=32$~meV in ML-WSe$_2$.\cite{Courtade_PRB17}  The following parameters were used for the Fr\"{o}hlich coupling [Eq.~(\ref{eq:Frohlich_coupling})]: The area of the unit cell is $A_u = \sqrt{3}a_{lc}^2/2 = 8.87~\AA^2$ where  $a_{lc}=3.2~\AA$ is the triangular lattice constant.  The atomic masses of tungsten and  selenium atoms are $M_{\text{W}}=3.05\cdot 10^{-22}$~g  and $M_{\text{Se}}=1.31\cdot 10^{-22}$~g. The Born effective charge is $Z_{E_2'}=-1.16$ for ML-WSe$_2$.\cite{Sohier_PRB16} For the calculation of the PL in ML-MoSe$_2$ (Fig.~\ref{fig:PL_B}), we have used the following respective parameters: $m_{cb} = 0.5m_0$,  $m_{ct} = 0.58m_0$, $m_{vt} = 0.6m_0$,\cite{Kormanyos_2DMater15}  $\alpha_P=0.5$,\cite{VanTuan_PRB18}  $E_{E_2'}=29.8$~meV,\cite{VanTuan_arXiv19} $M_{\text{Mo}}=1.59\cdot 10^{-22}$~g, and $Z_{E_2'}=-1.78$.\cite{Sohier_PRB16} 

\section{Benchmarking the SVM results against empirical findings}\label {app:benchmark_delocalized}

Table~\ref{tab:E_deloc} shows the calculated results for the ground-state energies of several delocalized few-body complexes in hBN-encapsulated ML-WSe$_2$ [assigning $V_{\text{E}}(r_{i},z)=0$ in Eq.~(\ref{eq:H})]. The calculated binding energies of the neutral exciton, $E_{X^0}=178$~meV, and biexciton, $\mathcal{E}_{XX^0}=17.4$~meV, can be compared directly  with emipirical results. The former agrees very well with the empirical  value found by magneto-absorption spectroscopy,~\cite{Stier_PRL18} and the latter is in excellent agreement with the empirical findings in Table~\ref{tab:EPosition}. 

The calculated binding energies of the dark exciton and trion states can be compared indirectly with experimental results. Starting with the dark exciton, its binding energy is larger than that of the bright exciton because of the heavier mass in the bottom valley of the conduction band compared with the mass in  the top valley, leading to $E_{X_D^0}=195.2$~meV vs $E_{X^0}=178$~meV. To benchmark the binding energy of the dark exciton against experiment, we employ Eqs.~(\ref{Bind_Bexciton}) and (\ref{Bind_Dexciton}) and write 
 \begin{eqnarray}
E_{X_D^0} - E_{X^0} =  \hbar \omega_{X^0}  - \hbar \omega_{X_D^0} - \Delta.
 \label{eq:Delta_DB}
\end{eqnarray}
Our calculations show that $E_{X_D^0} - E_{X^0} =17.2$~meV. The experiments show that $\hbar \omega_{X^0}  - \hbar \omega_{X_D^0}$ varies between 40 and 43 meV.\cite{Chen_NatComm18,Ye_NatComm18,Li_NatComm18,Barbone_NatComm18} By substituting these results in Eq.~(\ref{eq:Delta_DB}), we can validate the calculated values if the spin-splitting energy of the conduction band, $\Delta$, ranges between 23 and 26 meV. Indeed, this value is in excellent agreement with the one found in independent experiments.\cite{Zhang_NatNano17} 

Finally, we examine the binding energy of the dark negative trion, $\mathcal{E}_{X_D^-}  = 25.6$~meV.  While we cannot benchmark this result directly with experiment, we can compare the binding energy we got for the bright negative trion in its singlet form with the one seen in experiments. The empirically found binding energy of the latter is 35~meV,\cite{Courtade_PRB17,Wang_NanoLett17} and it is in excellent agreement with the calculated value of our model, $\mathcal{E}_{X^-}  = 34.9$~meV.\cite{VanTuan_PRB18}  We notice that the calculated binding energy of the negative bright trion is nearly 10~meV larger than that of the dark one (34.9 vs 25.6~meV). The reason is that the binding energy of a bright trion is taken with respect to a bright exciton whose electron has an effective mass $0.29m_0$ (top valley of the conduction band in ML-WSe$_2$). The bright negative trion is formed by adding an electron whose effective mass is heavier by $\sim$30\% (bottom valley of the conduction band in ML-WSe$_2$). As a result, the binding with respect to the bright exciton is tighter because we add a heavier electron to the complex. In the case of the dark trion, the binding energy is taken with respect to a dark exciton whose electron has an effective mass $0.4m_0$. The dark negative trion is formed by adding an electron with the same mass (from the time-reversed valley). In this case, there is no extra enhancement in the binding energy. In this view, the binding energy of the dark negative trion is closer to that of the bright positive one whose empirical value is 21~meV,\cite{Courtade_PRB17} and calculated value is 23.6~meV.\cite{VanTuan_PRB18} In this case, the effective mass of both holes in the time-reversed top valleys of the valence band is $0.36m_0$.

\end{document}